\input epsf

\newcommand{\bea}{\begin{eqnarray}}
\newcommand{\eea}{\end{eqnarray}}
\newcommand{\ba}{\begin{array}}
\newcommand{\ea}{\end{array}}
\newcommand{\be}{\begin{equation}}
\newcommand{\ee}{\end{equation}}

\newcommand{\Eqref}[1]{Eq.(\ref{#1})}
\newcommand{\eqref}[1]{eq.(\ref{eqn:\LBL:#1})}

\newcommand{\figcap}[1]{\addtocounter{figure}{1}
	{\bf Figure \thefigure}: {\small\sl #1}}
\newcommand{\tabcap}[1]{\addtocounter{table}{1}
	{\bf Table \thetable}: {\small\sl #1}}

\newcommand{\cL}{{\cal L}}
\newcommand{\cO}{{\cal O}}
\newcommand{\minus}{\!-\!}
\newcommand{\plus}{\!+\!}
\newcommand{\non}{\nonumber}
\newcommand{\inv}[1]{\frac{1}{#1}}
\def\lsim{\; \raise0.3ex\hbox{$<$\kern-0.75em
      \raise-1.1ex\hbox{$\sim$}}\; }
\def\gsim{\; \raise0.3ex\hbox{$>$\kern-0.75em
      \raise-1.1ex\hbox{$\sim$}}\; }

\newcommand{\abs}[1]{\left|#1\right|}

\newcommand{\im}{{\rm Im}}
\newcommand{\re}{{\rm Re}}
\newcommand{\bfT}{{\bf\rm\bf T}}
\newcommand{\Tr}{{\rm Tr\,}}
\newcommand{\del}{\partial}
\newcommand{\ra}{\rangle}
\newcommand{\la}{\langle}

\newcommand{\align}{\!\!\!\!&&}

\def\id{\leavevmode\hbox{\small1\kern-3.3pt\normalsize1}}

\newcommand{\rvacb}[1]{|{\cal O}_\beta #1\rangle}
\newcommand{\lvacb}[1]{\langle{\cal O}_\beta #1 |}

\newcommand{\annp}[3]{Ann.~Phys.~(N.Y.) {{\bf #1} {(#2)} {#3}}}
\newcommand{\apj}[3]{Ap.~J.~{{\bf #1} {(#2)} {#3}}}
\newcommand{\apjl}[3]{Ap.~J.~Lett.~{{\bf #1} {(#2)} {#3}}}

\newcommand{\app}[3]{Astropart.~Phys.~{{\bf #1} {(#2)} {#3}}}

\newcommand{\pr}[3]{Phys.~Rev.~{{\bf #1} {(#2)} {#3}}}

\newcommand{\prl}[3]{Phys.~Rev.~Lett.~{{\bf #1} {(#2)} {#3}}}

\newcommand{\pl}[3]{Phys.~Lett.~{{\bf #1} {(#2)} {#3}}}
\newcommand{\prep}[3]{Phys.~Rep.~{{\bf #1} {(#2)} {#3}}}
\newcommand{\ptp}[3]{Prog.~Theor.~Phys.~{{\bf #1} {(#2)} {#3}}}

\newcommand{\np}[3]{Nucl.~Phys.~{{\bf #1} {(#2)} {#3}}}

\newcommand{\jpg}[3]{J.~Phys.~G:~Nucl.~Part.~Phys.~{{\bf #1} {(#2)} {#3}}}

\newcommand{\zp}[3]{Z.~Phys.~{{\bf #1} {(#2)} {#3}}}

\newcommand{\spjetp}[3]{ Sov.\ Phys.\ JETP\ {{\bf #1} {(#2)} {#3}}}

\newcommand{\lbmeff}{\cL_{eff}^{\beta,\mu}}
\newcommand{\lbmeffzero}{\cL_{0}^{\beta,\mu}}
\newcommand{\lbmeffone}{\cL_{1}^{\beta,\mu}}
\newcommand{\lbmeffreg}{{\cL_{1,reg}^{\beta,\mu}}}
\newcommand{\lbmeffosc}{{\cL_{1,osc}^{\beta,\mu}}}

\newcommand{\AUTHORS}{\
{\centering
{\large Per Elmfors,\footnote{Email address:
elmfors@nordita.dk.}}\raisebox{1ex}{,a}
{\large David Persson\footnote{Email address:
tfedp@fy.chalmers.se.}}\raisebox{1ex}{,b}  and
{\large Bo-Sture Skagerstam\footnote{Email address:
tfebss@fy.chalmers.se. Research supported by the Swedish National
Research Council under contract no. 8244-311.}}\raisebox{1ex}{,b,c}
 \\
{\sl \raisebox{1ex}{a}NORDITA,
   Blegdamsvej 17,
 DK-2100 Copenhagen \O, Denmark \\ }
{\sl \raisebox{1ex}{b}Institute of Theoretical Physics,
   Chalmers University of Technology and \\
    University of G\"oteborg, S-412 96 G\"oteborg,
Sweden \\ }
{\sl \raisebox{1ex}{c}University of Kalmar,
Box 905, S-391 29 Kalmar, Sweden\\}
}}
\documentstyle [12pt]{article}

\begin{document}
\thispagestyle{empty}
\begin{flushright}
	NORDITA-93/78 P\\
        G\"{o}teborg ITP 93-11 \\
	hep-ph/9312226\\
	December 1993\\
 \end{flushright}
\begin{center}
\baselineskip 1.2cm
{\Large\bf   Real-Time Thermal Propagators\\
 and the QED Effective Action\\
  for an External  Magnetic Field}
\normalsize
\end{center}
%
\AUTHORS
%
\vfill
\begin{abstract}
{\normalsize
The thermal averaged real-time propagator  of a Dirac fermion in
a static uniform
magnetic field $B$ is derived. At non-zero chemical
potential and temperature  we find explicitly the effective
action for the magnetic field,
which is shown to be closely related to the Helmholz free energy
of a relativistic fermion gas, and it exhibits the expected
de\ Haas -- van\ Alphen oscillations.
An effective QED coupling constant at finite temperature
and density is derived,
and compared with renormalization group results.
We discuss some
 astrophysical implications of our results.
}  \end{abstract}
\newpage
\setcounter{page}{1}
\begin{center}
\section{\sc Introduction}
\label{introduction}
\end{center}
Large magnetic fields $B$ can  be associated with certain compact astrophysical
objects like  supernovae \cite{ginzburg91,myller90} where $B={\cal O}
(10^{10})$T, neutron stars \cite{neutronstar,Chanm92} where $B={\cal O}
(10^{8})$T, or white magnetic dwarfs \cite{Chanm92,angel78}
 in which case $B={\cal O}
(10^{4})$T. (As a reference the electron mass in units of Tesla is
$m^2_e/e = 4.414 \cdot 10^9$T.)
It has recently  been argued that a plasma at thermal equilibrium can
sustain fluctuations
of the electromagnetic fields. In particular, for the primordial
Big-Bang plasma the
amplitude of magnetic field fluctuations at the time of the
primordial nucleosynthesis can be as large as
$B = {\cal O}(10^{10}) $T \cite{tajima}. Furthermore, a model
for extragalactic gamma
bursts in terms of mergers of massive binary stars suggests magnetic
fields up to the order $B={\cal O}(10^{13})$T \cite{nar&pac&pir92}.
A more speculative system where even larger macroscopic magnetic fields can be
contemplated are superconducting strings \cite{wittenetc}.
Here one may conceive fields
as large as $B \gsim {\cal O}(10^{14})$T. It has also recently been suggested
that due to gradients in the Higgs
field during the electroweak phase
transition in the early universe very large magnetic fields,
$B = {\cal O}(10^{19})$T,
may be
generated \cite{vach91}. If one encounters magnetic fields of this order of
magnitude
the complete electroweak model has to be considered and the concept of
electroweak
magnetism becomes important (for a recent account see e.g.
Ref.\cite{ambj&ole93}). In
the present paper we  consider, however, magnetic fields such that calculations
within QED  are sufficient. A shorter version of this report has been published
elsewhere \cite{elm&per&ska93a}.

In many of these systems one has to consider the effects of a
thermal environment
 and a finite chemical potential. In this paper we derive the
appropriate effective fermion propagator and the effective action
in   QED   for a thermal
environment treated exactly in the external constant magnetic field
but with no virtual photons present, i.e. we consider the weak coupling limit.

Calculations of the QED effective Lagrangian density
 in an external field have been
attempted  before
either at finite temperature \cite{dittrich79,rojas92} or at
finite chemical potential \cite{ceo90}.  In
the latter case \cite{ceo90} the effective action is not complete but the
correct form is
presented here. At finite chemical potential and for sufficiently small
temperatures,
the QED effective action should exhibit a certain periodic dependence of
the
external field, i.e. the well-known de\ Haas -- van\ Alphen
oscillations in condensed
matter
physics. This was not obtained in Ref.\cite{ceo90}.
Elsewhere, the radiative corrections to the anomalous magnetic moment
has been estimated in the presence of large magnetic fields and it was
argued that they are extremely small \cite{pebss91,Studenikin90}.

 By making use of the effective action we derive
the effective QED coupling as a function of the
external field, the chemical potential and temperature.
In a future publication we
will
discuss the fermion self-energy and
radiative corrections to the electrons anomalous magnetic moment,
in terms of the formalism derived here
\cite{elm&per&ska93b}.

\begin{center}
\section{\sc Thermal propagators in the Furry picture}
\label{furry}
\setcounter{equation}{0} \end{center}
We consider Dirac fermions in the presence of an external static field as
described by the vector potential $A_\mu$. Using static energy solutions
we may represent  the second quantized
fermion field in the Furry picture \cite{furry51}. It is given by
\be
  \Psi({\bf x},t) = \sum _{\lambda,\kappa}
 b_{\lambda\kappa}\psi^{(+)}_{\lambda\kappa}({\bf x},t) +
  d_{\lambda\kappa}^{\dagger}\psi^{(-)}_{\lambda\kappa}({\bf x},t)~~~,
\ee
where $\lambda$ is a polarization index, $\kappa $ denotes the energy
 and momentum (or other) quantum numbers
(discrete and/or continuous) needed in order to completely characterize the
solutions, and
 $(\pm)$ denotes positive and negative energy solutions of
the corresponding Dirac equation,
\be
 (i\not\!\!D - m)\psi^{(\pm)}_{\lambda\kappa} ({\bf x},t) =
0~~~,
\ee
where $D_\mu = \partial_\mu +i eA_\mu$ is the covariant derivative.
 The creation and annihilation operators satisfy the canonical anti-commutation
relations
\be
\{d_{\lambda'\kappa '},d_{\lambda\kappa}^{\dagger}\} =
\delta_{\lambda'\lambda}\delta_{\kappa'\kappa} = \{b_{\lambda'\kappa
'},b_{\lambda\kappa}^{\dagger}\}~~~,
\ee
while other anti-commutators are zero. The completeness relation
\be
\sum _{\lambda,\kappa}
\psi^{(+)\dagger}_{\lambda\kappa,a}({\bf x'},t)\psi^{(+)}_{\lambda\kappa,b}
({\bf x},t) +  \psi^{(-)\dagger}_{\lambda\kappa,a}({\bf
x'},t)\psi^{(-)}_{\lambda\kappa,b}({\bf x},t) = \delta _{ab}
 \delta^3 ({\bf x'} -{\bf x})~~~,
\ee
where $\psi^{(\pm)}_{\lambda\kappa,a}$ denotes the $a$-component of the Dirac
 spinor $\psi^{(\pm)}_{\lambda\kappa}$,
leads to the canonical anti-commutation
relations for the fields
\be
\{ \Psi_{a}({\bf x'},t), \Psi^{\dagger}_{b}({\bf x},t)  \} = \delta
_{ab}\delta ^{3}({\bf x'} - {\bf x})~~~.
 \ee
In   vacuum, the
fermion propagator $iS_{F}(x';x|m)$ is defined by
\bea
iS_{F}(x';x|m)~=~\la 0|{\bf T}\left( \Psi({\bf x'},t') \overline{\Psi}({\bf
x},t)\right)|0\ra  =~~~~~~~~~~~~~~~~~~~~~~&&\nonumber \\
 \theta(t'-t)
\sum_{\lambda\kappa}\psi^{(+)}_{\lambda\kappa}({\bf x'},t')
\overline{\psi}^{(+)}_{\lambda\kappa}({\bf x},t) -  \theta(t-t')
\sum_{\lambda\kappa}\psi^{(-)}_{\lambda\kappa}({\bf x'},t')
\overline{\psi}^{(-)}_{\lambda\kappa}({\bf x},t)~~~,
\label{eq:vprop}
 \eea
where the conjugated spinor $\overline{\psi}^{(\pm)}_{\lambda\kappa}$is given
 by
 $\overline{\psi}^{(\pm)}_{\lambda\kappa} =
({\psi}^{(\pm)}_{\lambda\kappa})^{\dagger}\gamma_{0}$. Since
  $\psi^{(\pm)}_{\lambda\kappa}({\bf x},t)$ satisfies the Dirac equation, only
the time derivative acting on the step functions   gives a non-zero
contribution, so one finds that
\be
 (i\not\!\!D - m)S_{F}(x';x|m) = \id \cdot\delta^{4}(x' - x)~~~.
\label{eq:delteq}
\ee

The  real-time  propagator at finite temperature $T$ and
chemical potential
$\mu$,  denoted by $\la iS_{F}(x';x|m)\ra_{\beta,\mu} $, can now be obtained by
the following
reasoning. Let $f^{+}_{F}(\omega )$ denote the Fermi-Dirac thermodynamical
distribution function
\be
\label{eq:fpmdistribution}
f^{+}_{F}(\omega ) = \frac{1}{\exp (\beta (\omega - \mu )) + 1}~~~,
\ee
where $\beta$ is the inverse temperature, $\mu $ the chemical potential and
$\omega$ is the energy of the quantum state under consideration. A particle can
propagate forward in time in a state which is
unoccupied by thermal particles, whereas a hole in the occupied states
 can propagate backwards in  time . We can therefore write
\bea
\la iS_{F}(x';x|m)\ra_{\beta,\mu}
& = &\sum_{\lambda,\kappa} \nonumber \\
  \left[\theta(t'-t)
\left([1-f^{+}_{F}(E_{\kappa})]\psi^{(+)}_{\lambda\kappa}({\bf x'},t')
\overline{\psi}^{(+)}_{\lambda\kappa}({\bf x},t) \right.\right.& +
&\left.\left.
[1-f^{+}_{F}(-E_{\kappa})]\psi^{(-)}_{\lambda\kappa}({\bf x'},t')
\overline{\psi}^{(-)}_{\lambda\kappa}({\bf x},t) \right)
 \right.  \nonumber \\
- \theta(t-t') \left.
\left(f^{+}_{F}(-E_{\kappa})\psi^{(-)}_{\lambda\kappa}({\bf x'},t')
\overline{\psi}^{(-)}_{\lambda\kappa}({\bf x},t) \right.\right.& +
&\left.\left.
f^{+}_{F}(E_{\kappa})\psi^{(+)}_{\lambda\kappa}({\bf x'},t')
\overline{\psi}^{(+)}_{\lambda\kappa}({\bf x},t) \right)
 \right]~.
\label{eq:time}
\eea
We can now  extract the vacuum part of the propagator Eq.(\ref{eq:vprop}) and
write
\be
      \la iS_{F}(x';x|m)\ra_{\beta,\mu}  =
      iS_{F}(x';x|m) + iS_{F}^{\beta,\mu}(x';x|m)~~~,
\label{eq:thermalS}
\ee
where  the thermal part $iS_{F}^{\beta,\mu}(x';x|m)$ is defined by
\bea
&&~~~~~~~~~~~~~~~~~~~~S_{F}^{\beta,\mu}(x';x|m)~~~=  \nonumber \\
&&i \sum_{\lambda,\kappa}
\left(f^{+}_{F}(E_{\kappa})\psi^{(+)}_{\lambda\kappa}({\bf x'},t')
\overline{\psi}^{(+)}_{\lambda\kappa}({\bf x},t)  -
f^{-}_{F}(E_{\kappa})\psi^{(-)}_{\lambda\kappa}({\bf x'},t')
\overline{\psi}^{(-)}_{\lambda\kappa}({\bf x},t) \right)~~~,
\label{eq:thermalpart}
\eea
and where we have defined the distribution
\be
\label{eq:fminus}
f^{-}_{F}(E_{\kappa}) = 1 - f^{+}_{F}(-E_{\kappa})~~~.
\ee
Notice  that there is no time-ordering in $S_{F}^{\beta,\mu}(x';x|m)$
 despite the fact that the
time-ordering in Eq.(\ref{eq:time}) is non-trivial. The
thermal propagator Eq.(\ref{eq:thermalS}) therefore also trivially satisfies
Eq.(\ref{eq:delteq}). These considerations can, of course, easily be extended
to
treat particles with Bose-Einstein statistics as well.

The result
Eq.(\ref{eq:thermalpart}) can also be derived from an
explicit calculation using the second-quantized field operators and appropriate
thermal averages, i.e. we use Wicks theorem
\be
{\bf T}\left( \Psi({\bf x'},t') \overline{\Psi}({\bf
x},t)\right) = iS_{F}(x';x|m)~+~
:\Psi({\bf x'},t') \overline{\Psi}({\bf
x},t):~~~,
\ee
where the last term corresponds to a normal ordering. We then obtain
\be
\la{\bf T}\left( \Psi({\bf x'},t') \overline{\Psi}({\bf
x},t)\right)\ra_{\beta,\mu}  = iS_{F}(x';x|m) + iS_{F}^{\beta,\mu}(x';x|m)~~~,
\ee
where we have used the only non-zero bilinear thermal
averages
\bea
\label{fermidirac}
      \la b_{\lambda\kappa}^{\dagger}b_{\lambda'\kappa'}\ra_{\beta,\mu}
      &=& f_{F}^{+}(E_{\kappa})\delta_{\lambda\lambda'}
       \delta_{\kappa\kappa'}~~~,\nonumber\\
      \la d_{\lambda\kappa}^{\dagger}d_{\lambda'\kappa'}\ra_{\beta,\mu}
      &=& f_{F}^{-}(E_{\kappa})\delta_{\lambda\lambda'}
      \delta_{\kappa\kappa'}~~~.
\eea
In principle we do not have to restrict ourselves to thermal distributions as
given by
\Eqref{eq:fpmdistribution}. In fact, we can allow for {\it any} such
one-particle
distribution function $f^{+}_{F}(\omega )$ and the definition
\Eqref{eq:fminus}.
\begin{center}
\section{\sc External uniform and static magnetic field}
\label{magnetic}
\setcounter{equation}{0}
\end{center}
For the convenience of the reader, we summarize some of the relevant
expressions in the
case of a constant magnetic field $B$ parallel to the $z-$direction in the
gauge $A_\mu=(0,0,-Bx,0)$. Using
$\kappa$ as
a collective index for $(n,k_{y},k_{z})$, where $n=0,1,2,...$  ; $k_{y},k_{z}$
are continuous, and the $\gamma$ matrices   in the chiral representation, we
can write the
 solutions in the form
\be
  \psi^{(\pm)}_{\lambda,\kappa}({\bf x},t)= \frac{1}{2\pi}\, \frac{\exp[ \pm (
  -i E_{\kappa}t \plus i k_{y}y \plus i k_{z}z ) ]}{ \sqrt{2 E_{\kappa} } } \,
    \Phi^{(\pm)}_{\lambda,\kappa }(x)~~~ ,
\ee
where
\be
  \Phi^{(+)}_{1,\kappa}(x) =\frac{1}{\sqrt{E_{\kappa}\plus k_{z} } }
    \left( \ba{c}
             (E_{\kappa} \plus k_{z} ) I_{n;k_{y}}(x) \\
             - i \sqrt{2eBn}\, I_{n-1;k_{y}}(x) \\
             - m I_{n;k_{y}}(x) \\
              0
          \ea \right)~~~ ,
\ee
\be
 \Phi^{(+)}_{2,\kappa}(x) =\frac{1}{\sqrt{E_{\kappa}\plus k_{z} } }
    \left( \ba{c}
             0 \\
             - m I_{n-1;k_{y}}(x) \\
             - i \sqrt{2eBn}\, I_{n;k_{y}}(x) \\
             ( E_{\kappa} \plus k_{z} ) I_{n-1;k_{y}} (x)
           \ea \right)~~~ ,
\ee
\be
   \Phi^{(-)}_{1,\kappa}(x) =\frac{1}{\sqrt{E_{\kappa}\minus k_{z} } }
    \left( \ba{c}
           -m I_{n;-k_{y}}(x) \\
           0 \\
           (-E_{\kappa} \plus k_{z} ) I_{n;-k_{y}}(x) \\
           i \sqrt{2eBn} \,  I_{n-1;-k_{y}}(x)
          \ea  \right)~~~ ,
\ee
\be
  \Phi^{(-)}_{2,\kappa}(x) =\frac{1}{\sqrt{E_{\kappa}\minus k_{z} } }
    \left( \ba{c}
           i \sqrt{2eBn} \,  I_{n;-k_{y}}(x) \\
            (-E_{\kappa} \plus k_{z} ) I_{n-1;-k_{y}}(x) \\
             0 \\
             - m  I_{n-1;-k_{y}}(x)
           \ea \right) ~~~ .
\ee
In these expressions the energy $E_{\kappa}$ is given by
\be
E_{\kappa} = \sqrt{m^{2} + k^{2}_{z} +2eBn}~~~,
\ee
and the $ I_{n;k_{y}}(x)$ functions  are explicitly written
\bea
  I_{n;k_{y}}(x)& \equiv& \left( \frac{eB}{\pi} \right)^{1/4} \exp \left[
  - \frac{1}{2} eB \left( x \minus \frac{k_{y}}{eB} \right)^{2} \right]
 \frac{1}{ \sqrt{n!}} H_{n} \left[ \sqrt{2eB} \left( x \minus \frac{k_{y}}
 {eB} \right) \right] ~~~ . \nonumber \\
 &&
\label{eq:Is}
\eea
Here  $H_{n}$ is the Hermite polynomial given by Rodrigues' formula as
\be
  H_{n}(x)=(-1)^{n} e^{\frac{1}{2}x^{2} } \frac{d^{n}}{dx^{n}} e^{- \frac{1}{2}
            x^{2}} ~~~ ,
\ee
and we define $ I_{-1;k_{y}}(x)=0$.
The functions $ I_{n;k_{y}}(x)$ are normalized as
\be
\label{Iid}
\int dx I_{n;k_{y}}(x) I_{m;k_{y}}(x) = \delta _{n,m}~~~
\ee
if $ n,m\geq 0$, so that it is easily shown that the collection of
all $\Psi$'s form a complete orthonormal set.
The vacuum part of the propagator Eq.(\ref{eq:vprop})
is then given by (see e.g. Ref.\cite{kobsak83})
\bea
   S_{F}(x';x|m)_{ab}&=& \sum^{\infty}_{n=0}\int
\frac{d\omega \, dk_{y} \, dk_{z}}{(2\pi)^{3}} \exp[-i\omega (t' -t) +
 ik_{y}(y' - y) + ik_{z}(z' - z)] \nonumber \\
 & & \times\frac{1}{\omega^{2}
 \minus k_{z}^{2} \minus m^{2} \minus 2eBn  +i\epsilon}
  S_{ab}(n;\omega , k_{y}, k_{z}) ~~~.
\label{bvprop}
\eea
 The matrix $S(n;\omega , k_{y}, k_{z})$ entering above  is given by
\pagebreak
\bea
  \lefteqn{~~~~~~~~~~~~~~~~~~~~~~~~~~~~~~~S(n;\omega , k_{y}, k_{z}) \equiv}
 \nonumber
\\[2ex]
 && \left( \ba{cccc}
     mI_{n,n} & 0 & - (\omega \plus k_{z}) I_{n,n} & - i \sqrt{2eBn}
                  I_{n,n-1} \\
    0 & m I_{n-1,n-1} & i \sqrt{2eBn} I_{n-1,n} & - (\omega \minus k_{z})
                  I_{n-1,n-1} \\
    - (\omega \minus k_{z} ) I_{n,n} & i \sqrt{2eBn} I_{n,n-1} &
       m I_{n,n} & 0 \\
    - i \sqrt{2eBn} I_{n-1,n} & - (\omega \plus k_{z} ) I_{n-1,n-1} &
       0 & m I_{n-1,n-1}
    \ea \right)~~~ ,\nonumber \\
\eea
where we have defined
\be
  I_{n,n'} \equiv I_{n;k_{y}}(x) I_{n';k_{y}}(x')~~~~.
\ee
Similarly we find the thermal part of the fermion propagator
\bea
S^{\beta,\mu}_{F}(x';x|m)_{ab}&=& \sum^{\infty}_{n=0}\int
\frac{d\omega \, dk_{y} \, dk_{z}}{(2\pi)^{3}} \exp[-i\omega (t' -t) +
 ik_{y}(y' - y) + ik_{z}(z' - z)] \nonumber \\
 & & \times 2  \pi i \, \delta( \omega^{2}
 \minus k_{z}^{2} \minus m^{2} \minus 2eBn )
 f_{F}(\omega)  S_{ab}(n;\omega , k_{y}, k_{z}) ~~~,
\label{btprop}
\eea
where  $f_{F}(\omega )$ is the thermal distribution
 \be
 f_{F}(\omega) = \theta(\omega )f_{F}^{+}(\omega)
 \plus \theta(-\omega ) f^{-}_{F}(-\omega)~~~.
\ee
By making use of the completeness relation
\be
\sum _{n=0}^{\infty} I_{n;k_{y}}(x) I_{n;k_{y}}(x') = \delta (x - x')\ ,
\ee
one can show \cite{elm&per&ska93b} that the propagators \Eqref{bvprop} and
\Eqref{btprop} reduce to the free-field propagators in the limit when the
magnetic field $B$ tends to
zero.
\setcounter{equation}{0}
\begin{center}
\section{\sc Propagators in thermo-field dynamics}
\label{thermalfieldpropagator}
\end{center}
The thermal propagators in Eqs.(\ref{bvprop}) and (\ref{btprop}) cannot be
used for a perturbative expansion in a naive way. The reason
is that the $\delta$-functions can occur on several
internal legs with coinciding arguments and that such expressions are  not
well-defined. It is known that such problems can be avoided
be means of a correctly derived
real-time finite temperature formalism where one must invoke a
doubling of the degrees of freedom. There are several formalisms
for doing that and we shall use thermo field dynamics (TFD) since
it is very easy in the operator formalism \cite{UmezawaMT82}.
In TFD the propagator is obtained as the expectation value
of the time-ordered product in the thermal vacuum $\rvacb{}$
which is annihilated by the thermal operators
$(b_{\lambda\kappa}(\beta),\ d_{\lambda\kappa}(\beta))$ and their
tilde partners $(\tilde{b}_{\lambda\kappa}(\beta),
\ \tilde{d}_{\lambda\kappa}(\beta))$.

The TFD propagator can  be given by a simple expression
for independent harmonic oscillators.
We have solved the
Dirac equation exactly in the external field, but in  the
free propagator the interaction between the particles is neglected.
Each mode is therefore still an independent harmonic
oscillator, but with a different frequency labeled by
the quantum numbers $(n,k_y,k_z)$ and corresponding to a
definite Landau level. Thus, in the derivation of the
propagator we can copy the usual procedure for
free particles.

The Bogoliubov transformation between the zero temperature and
thermal operators is given by
\bea
\label{Btrf1}
\left(\ba{c} b_{\lambda\kappa}\\ i\tilde{b}^\dagger_{\lambda\kappa} \ea\right)
&=&
\left(\ba{cc} \cos\vartheta^{(+)}_{\lambda\kappa} &
      -\sin\vartheta^{(+)}_{\lambda\kappa}\\
      \sin\vartheta^{(+)}_{\lambda\kappa} &
      \cos\vartheta^{(+)}_{\lambda\kappa} \ea\right)
\left(\ba{c} b_{\lambda\kappa}(\beta)\\
i\tilde{b}^\dagger_{\lambda\kappa}(\beta) \ea\right)   \ ,
\eea
and
\bea
\label{Btrf2}
\left(\ba{c} d_{\lambda\kappa}\\ i\tilde{d}^\dagger_{\lambda\kappa} \ea\right)
&=&
\left(\ba{cc} \cos\vartheta^{(-)}_{\lambda\kappa} &
 -\sin\vartheta^{(-)}_{\lambda\kappa}\\
      \sin\vartheta^{(-)}_{\lambda\kappa} &
\cos\vartheta^{(-)}_{\lambda\kappa} \ea\right)
\left(\ba{c} d_{\lambda\kappa}(\beta)\\
i\tilde{d}^\dagger_{\lambda\kappa}(\beta) \ea\right) \ .
\eea
The number expectation values in Eq.(\ref{fermidirac}) imply that the
coefficients in the Bogoliubov matrices must satisfy
\be
	\sin^2\vartheta^{(\pm)}_{\lambda\kappa}=
	f^\pm_F(E_\kappa)\ .
\ee
We use the convention that  $b$ and $\tilde{b}$ anti-commute.
The Bogoliubov matrices in Eqs.(\ref{Btrf1}) and (\ref{Btrf2}), and the factors
of $i$, are carefully explained in Ref.\cite{Ojima81}. The definition
of the fermion propagator varies slightly in the literature
and we shall for definiteness follow Ref.\cite{Ojima81}.
Other propagators correspond to other definitions of
the thermal doublet and they may be computed in a
similar way. Since we are not doing any higher loop
calculations these conventions do not matter.
We compute the TFD
propagator matrix as
\be
iS_{F}^{TFD}(x';x|m)_{ab}=
\lvacb{} \bfT \left[ \left(\ba{c}\Psi_a(x') \\
      i\tilde{\Psi}_a^\dagger(x') \ea\right)
      \left(\overline{\Psi}_b(x),\ \
      -i\tilde{\overline{\Psi}}_b^\dagger(x)\right)\right]
\rvacb{}\ .
\ee
The structure of the propagator is the same as in absence of the
external field except that we now expand in another basis
corresponding to the new energy eigenvalues. We obtain
 \bea
iS_{F}^{TFD}(x';x|m)_{ab}&=&
\sum^{\infty}_{n=0}\int
\frac{d\omega \, dk_{y} \, dk_{z}}{(2\pi)^{3}} \exp[-i\omega (t'-t) +
 ik_{y}(y'- y) + ik_{z}(z'-z)] \non\\
&&\times S_{ab}(n;\omega,k_y,k_z) U_F(\omega)
\left(\ba{cc} \inv{\omega^2-E^2_\kappa+i\epsilon} & 0 \\
0 & \inv{\omega^2-E^2_\kappa-i\epsilon} \ea \right) U^T_F(\omega)\ ,
\eea
where
\bea
U_F(\omega)&=&\left(\ba{cc} \cos\vartheta(\omega) &
-\sin\vartheta(\omega)\\
\sin\vartheta(\omega) & \cos\vartheta(\omega) \ea\right)\ ,
\eea
and
\bea
\sin\vartheta(\omega)&=&\theta(\omega)\sqrt{f^+_F(\omega)}
      -\theta(-\omega)\sqrt{f^-_F(-\omega)}\ ,
\non \\
\cos\vartheta(\omega)&=&\theta(\omega)\sqrt{1-f^+_F(\omega)}
      +\theta(-\omega)\sqrt{1-f^-_F(-\omega)}\ .
\eea
Here, $U^T_F(\omega)$ is the transpose of the matrix $U_F(\omega)$.
The $S^{TFD}_F(x';x|m)_{ab}^{11}$ component is, of course, the same
as the propagator in Eqs.(\ref{bvprop}) and (\ref{btprop}) and the other
components are only needed in higher loop calculations.

The derivation of the propagator in this Section can be repeated for a
non-equilibrium distribution if only we assume certain factorization properties
of the density matrix. The essential assumption is that there are no
non-trivial multiparticle correlations so that everything is determined in
terms of the single particle distribution. This freedom amounts to replacing
the functions $f^\pm_F(E_\kappa)$ with some other positive functions that
describes the distribution. Some applications of such a formalism in absence of
the external field can be found in Ref.\cite{ElmforsEV93c}.
%
\setcounter{equation}{0}
\begin{center}
\section{\sc The effective action}
\label{effective}
\end{center}
As was shown by Schwinger a long time ago \cite{Schwinger51}, an
external electromagnetic field, slowly varying in space and time,
can be treated to all orders in the external field
in the weak-coupling limit. Here we make use of
a technique similar to that of Schwinger's in order
to evaluate the thermodynamical partition function in a static uniform
magnetic field $B$ for charged fermions as well as
 for charged bosons.\vspace{4ex}

\subsection{\sc QED and  Charged Fermions}
The generating functional of fermionic Green's functions in an external
field $Z[\bar{\eta},\eta, A_{\mu}]$, formally defined by
\be
Z[\bar{\eta},\eta, A_{\mu}] = \int d[\bar{\psi}]d[\psi]
\exp[i\int d^{4}x(-\frac{1}{4}F_{\mu \nu}F^{\mu \nu} +
\bar{\psi}(i\not\!\!D - m)\psi - \bar{\eta}\psi + \bar{\psi}\eta)]~~,
\label{eq:genfunc}
\ee
describes second-quantized electrons and positrons  interacting with a
classical electromagnetic field expressed in terms of the
vector potential $A_{\mu}$. The expectation value of $\psi$ (and
$\bar{\psi}$) can formally be fixed by choosing appropriate $\bar{\eta}$ (and
$\eta$), i.e. $\varphi (x) \equiv \la\psi (x)\ra  =
-i\delta/\delta\bar{\eta}(x)\log Z$ (and similarly for $\bar{\psi}$). The
equation of motion for $\varphi (x)$ tells us how the electrons interact
with the electromagnetic field which includes effects due to all virtual
$e^{+}e^{-}$-pairs.

The fermionic Gaussian functional integral in Eq.(\ref{eq:genfunc}) can
formally be performed with
the
result that
\bea
\lefteqn{Z[\bar{\eta},\eta, A_{\mu}] =}  \nonumber \\
&&\mbox{Det}\left[i(i\not\!\!D - m)\right]\exp\left[i\int d^{4}x
\left( - \frac{1}{4}F_{\mu \nu}F^{\mu \nu}
\plus \int d^{4}y \bar{\eta}(x) S_{F}(x;y|m)\eta (y)\right)\right]~~~,
\eea
where $S_{F}(x;y|m)$ is the external field vacuum propagator as given by
Eq.(\ref{bvprop}). It follows that $\varphi (x)$ satisfies the Dirac
equation in the external field, i.e. $(i\not\!\!D -m)\varphi (x) =0$.

The functional determinant $\mbox{Det}(i\not\!\!D -m)$ gives rise
to a contribution  to  the effective  Lagrangian density
${\cal L}_{eff}$. Using a complete orthogonal basis to rewrite $\log
\mbox{Det}$ as $\Tr \log$,
the effective action can thus be written
\be
\label{Seff}
S_{eff} = \int d^{4}x {\cal L }_{eff} =  \int d^{4}x
\left[ -\frac{1}{4}F_{\mu \nu}F^{\mu \nu}\right]  \minus i \,  \Tr \log
 \left[i(i\not\!\!D -m) \right]~~~.
\ee
We now write the effective Lagrangian density as
\be
 {\cal L }_{eff} ={\cal L}_{0} \plus {\cal L}_{1}~~~,
\ee
where the tree level part in the case of a pure magnetic field is
\be
{\cal L}_{0}= -\frac{1}{2} B^{2}~~~,
\ee
and $ {\cal L}_{1}$ corresponds to the functional determinant.
Differentiating \Eqref{Seff} with
 respect to the fermion mass we now find the one-loop correction according to
\be
\label{dldm}
\frac{\partial  {\cal L}_{1}}{\partial m} = i \,{\rm tr} S_{F}(x;x|m)~~~,
\ee
where the trace now only is over spinor indices.
After a straightforward calculation of the trace
using Eq.(\ref{bvprop}), we obtain in terms of renormalized quantities
 the well-known
result \cite{Schwinger51} that
\begin{equation}
\label{TzeroEA}
{\cal L}_{1} = -\frac{1}{8\pi^{2}} \int_{0}^{\infty}
 \frac{ds}{s^3} \biggl[esB\coth(esB) -1 -
\frac{1}{3}(esB)^2\biggr]\exp(-m^{2}s)~~~.
\end{equation}
We have here  performed the standard
renormalizations leaving $eB$ invariant, i.e.
\bea
 A_{\mu} &\longrightarrow & (1+ Ce^{2})^{-1/2}A_{\mu}~~~, \nonumber \\
e^{2} &\longrightarrow & e^{2}\left(1+ Ce^{2}\right)~~~,
\label{chargeren}
\eea
where the divergent constant $C$ is given by
\be
C =\frac{1}{12\pi^{2}} \int_{0}^{\infty}
 \frac{ds}{s}\exp(-m^{2}s)~~~.
\ee

We shall now find the corresponding correction
 $S_{eff}^{\beta,\mu} = \int d^{4}x {\cal L}_{eff}^{\beta,\mu}~,$
to the effective action $S_{eff}$ at
finite chemical potential and temperature such that
\be
 \cL_{eff}=\cL_{0} \plus \cL_{1} \plus \lbmeff~~~.
\ee
We notice that
the correction ${\cal L}_{eff}^{\beta,\mu}$,
 due to the presence of  thermal
fermions, can be written in the form
\be
\frac{\partial {\cal L}_{eff}^{\beta,\mu}}{\partial m} =
i\mbox{Tr}S_{F}^{\beta,\mu}(x;x|m)~~~.
\label{thermaltrace}
\ee
By performing the trace operation in Eq.(\ref{thermaltrace}), using the
thermal propagator Eq.(\ref{btprop}), we obtain
\bea
 {\cal L}_{eff}^{\beta,\mu} &=&
\frac{4eB}{(2\pi )^{2}} \sum _{n=0}^{\infty}
\sum _{\lambda=1}^{2} \int _{-\infty}^{\infty}
d\omega f_{F}(\omega) \nonumber \\
&& \times\int _{0}^{\infty}
dk k^{2}  \delta (\omega ^{2}- k^{2} -2eB(n+\lambda-1) - m^{2})~~~,
\eea
where we have integrated by parts  with respect to $k$.
 We, therefore, see that ${\cal L}_{eff}^{\beta,\mu}$
is directly related to the partition function $Z(B,T,\mu)$ of
the relativistic fermion gas in the presence of an external
magnetic field $B$ in a sufficiently large quantization volume $V$,
as given in for example Ref.\cite{MillerR84},
 according to
\bea
\label{partition}
  \lbmeff &=& \frac{\log Z(B,T,\mu)}{\beta V}
 \nonumber \\
&=& \frac{eB}{(2\pi)^{2}}\sum _{n=0}^{\infty}
\sum _{\lambda=1}^{2} \int _{-\infty}^{\infty} dk  \frac{k^{2}}{E_{\lambda,n}}
 \nonumber \\
&& \times \left( \frac{1}{1+\exp[\beta(E_{\lambda,n} -\mu )]} +
  \frac{1}{1+\exp[\beta(E_{\lambda,n} +\mu )]}\right)~~~,
\eea
where
\be
E_{\lambda,n} = \sqrt{ k^{2} + 2eB(n+\lambda -1) + m^{2}}~~~.
\ee
Separating the field independent part we write
\be
{\cal L}_{eff}^{\beta,\mu} = {\cal L}_{0}^{\beta,\mu} +
{\cal L}_{1}^{\beta,\mu}~~~,
\ee
where
\be
\label{Lbmnoll}
      {\cal L}_{0}^{\beta,\mu} =
      \frac{1}{3\pi ^{2}} \int _{-\infty}^{\infty}
      d\omega  \theta(\omega ^{2} - m^{2})
      f_{F}(\omega )\left(\omega ^{2} - m^{2}\right)^{3/2}~~~.
\ee
We, therefore, conclude that the field independent thermal correction
 to the Lagrangian density $ {\cal L}_{0}^{\beta,\mu}$ can be identified as
\be
      {\cal L}_{0}^{\beta,\mu} = \frac{\log Z(T,\mu)}{\beta V}
      =-\frac{F(T,\mu)}{V}~~~,
\label{lbmeffzero}
\ee
where $ Z(T,\mu)$ is the partition function, and $F(T,\mu)$ the free
energy, for an ideal $e^{+}e^{-}$-gas
with particle energy $E=\sqrt{k^{2} + m^{2}}$, i.e.
\be
\frac{\log Z(T,\mu)}{V} = 2\int \frac{d^{3}k}{(2\pi)^{3}}
\left(\log[1+e^{-\beta(E-\mu)}] +
\log[1+e^{-\beta(E+\mu)}]\right)~~~,
\ee
consistent with the general identification above.
Using the identity
\be
\frac{\exp(-|x|)}{|x|} = \int _{0}^{\infty}
\frac{dt}{\sqrt{2\pi t}}\exp \left(
-\frac{1}{2}(x^{2}t +\frac{1}{t})\right)~~~,
\ee
the following  representation of ${\cal L}_{1}^{\beta,\mu}$, valid
 for $|\mu| < m$, can be derived in a straightforward manner
\be
\label{dittricheqn}
 {\cal L}_{1}^{\beta,\mu} =
  \frac{1}{4 \pi ^{2}}\sum _{l=1}^{\infty}(-1)^{l+1}
\int _{0}^{\infty} \frac{ds}{s^{3}}
\exp\left( -\frac{\beta ^{2} l^{2}}{4s} -m^{2}s \right)
\frac{\cosh(\beta l\mu)}{2}[eBs\coth(eBs) - 1]~~~.
\ee
In the case $\mu = 0$,
Eq.(\ref{dittricheqn}) agrees with
the result obtained in Refs.\cite{dittrich79,rojas92}.
However, it
is not always obvious, when written in this form, to see how
to extract the physical
content, and particularly not obvious how to generalize $\lbmeff$
to $\abs{\mu}\geq m$, since then it appears to be divergent.
In particular we notice that the high $T$ behaviour given in
Ref.\cite{dittrich79} is not correct. As explained
in Appendix A it is, however, possible to show that
 Eq.(\ref{dittricheqn}) is equal to \Eqref{Lbmueff}
given below, which is valid for all
$T$ and $\mu$.

In order to calculate the thermal part $\lbmeff$ of the effective action
in a more useful form, we have to be
careful with the convergence and the analytical structure.
Some details of the calculation are given in  Appendix A. We
get $\lbmeff = \lbmeffzero +\lbmeffone$, where
$\lbmeffzero$, the ideal gas contribution in absence of the
external field $B$, is
given in  \Eqref{lbmeffzero}, and
\bea
\lbmeffone &=& \lbmeffreg + \lbmeffosc \non\\
&=&
      \int_{-\infty}^\infty d\omega
      \theta(\omega^2-m^2)f_F(\omega)
      \Biggl[\inv{4\pi^{5/2}}\int_0^\infty
      \frac{ds}{s^{5/2}}e^{-s(\omega^2-m^2)}
      [seB\coth (seB) -1]\Biggr]\non\\
 &-&
\label{Lbmueff}
      \!\!\!\int_{-\infty}^\infty d\omega
      \theta(\omega^2-m^2)f_F(\omega)
      \Biggl[
      \inv{2\pi^3}\sum_{n=1}^\infty \left(\frac{eB}{n}\right)^{3/2}
      \sin \! \left(\frac{\pi}{4}-\frac{\pi n}{eB}
      (\omega^2-m^2)\right) \Biggr]\ .
\eea
The term with the
sum over $n$, $ \lbmeffosc$, was neglected in Ref.\cite{ceo90} and we
show in Section~\ref{physical}  that it is essential to keep
this term in order to get the
correct physical result. We may also use the generalized
$\zeta$-function to
rewrite $\lbmeffosc$ in a different  form, sometimes more
 suited for numerical calculations
\be
\label{eqovan}
\lbmeffosc = \int_{-\infty}^\infty d\omega
      \theta(\omega^{2}-m^{2})f_F(\omega) (eB)^{3/2}\frac{ \sqrt{2}}{\pi^{2}}
      \zeta \! \left( -\frac{1}{2},\mbox{ mod}\! \left[ \frac{ \omega^{2}
-m^{2}}{2eB} \right] \right)~~~,
\ee
where $\mbox{mod}[A]$ is a shorthand notation for $A$ modulo $1$, i.e.
\be
\mbox{mod}[A]= A-  \mbox{ int}[A]~~~.
\ee
An alternative  way to write \Eqref{eqovan} is
\bea
\lbmeffosc& =& \sum_{n=0}^{\infty} \int_{0}^{1} \frac{ds}
{\sqrt{m^{2} + 2eB(n+s)}}  \non \\
   && \times\left(f^{+}_{F}(\sqrt{m^{2} + 2eB(n+s)}) +
 f^{-}_{F}(\sqrt{m^{2} + 2eB(n+s)})\right) \zeta ( -\frac{1}{2},s)\ ,
\eea
where the various Landau-level contributions are made explicit.
In addition to $\lbmeff$ the free energy has a contribution
from the thermal photons, i.e.
\be
      \frac{F_\gamma(T)}{V} = -\frac{T^4\pi^2}{45}\ ,
\ee
which is background field independent since there is  no
self-interaction among abelian gauge fields.
%
\subsection{\sc QED and  Charged Scalars}
\label{scalarqed}
The formalism used so far applies also to scalar QED. We give some of the
corresponding results here for completeness. Equation (\ref{dldm}) becomes in
this case
\be
	\frac{\partial  {\cal L}_{1}}{\partial m^2} = -iG_{F}(x;x|m^2)~~~,
\ee
and the thermal propagator is
\bea
\la	G_F(x;x|m^2)\ra_{\beta,\mu}&=&\sum_{n=0}^\infty
	\int\frac{d\omega dk_y dk_z}{(2\pi)^3}
	\left(I_{n;k_y}(x)\right)^2  \non\\
	&&\times\left[\inv{\omega^2-E_n^2+i\epsilon}
	-2\pi i\delta(\omega^2-E_n^2)
	f_B(\omega)\right]\ ,
\eea
where
\be
	E_n^2 = k_z^2+(2n+1)eB+m^2\ ,
\ee
and
\be
	f_B(\omega) = \frac{\theta(\omega)}{e^{\beta(\omega-\mu)}-1}+
	\frac{\theta(-\omega)}{e^{\beta(-\omega+\mu)}-1}\ .
\ee
It is rather straightforward to obtain the correction
\be
	\cL_1 = \inv{16\pi^2}\int_0^\infty\frac{ds}{s^3}
	\exp (-m^2 s)\left(\frac{eBs}{\sinh(eBs)}-1+
	\frac{(eBs)^2}{6}\right)\ ,
\ee
to the effective action in the vacuum sector. At finite chemical
potential and temperature we similarly find the following contribution to
the effective action
\bea
	\lbmeff\align=\inv{6\pi^2}\int d\omega\theta(\omega^2-m^2-eB)
	f_B(\omega)(\omega^2-m^2)^{3/2}\non\\
	\align+\int d\omega\theta(\omega^2-m^2-eB)f_B(\omega)
	\left[\inv{8\pi^{5/2}}\int\frac{ds}{s^{5/2}}
	e^{-s(\omega^2-m^2) } \left(\frac{eBs}{\sinh(eBs)}-1\right)\right]\non\\
	\align-\int d\omega\theta(\omega^2-m^2-eB)f_B(\omega)
	\left[\inv{4\pi^3}\sum_{k=1}^\infty\left(\frac{eB}{k}\right)^{3/2}
	\sin\left(\frac{\pi}{4}-\frac{\pi k}{eB}
	(\omega^2-m^2-eB)\right)\right]\ .\non\\
\eea
The zero temperature part $\cL_1$ was derived in Ref.\cite{Schwinger51}.
Physically
this effective action is quite different from the fermionic one. We shall not
pursue this investigation here but only make a few remarks. Since for charged
bosons there is no
sharp Fermi surface, there are no de Haas -- van Alphen oscillations either.
Furthermore, even the
energy of the lowest Landau level ($n=0$) depends on $B$, so that,
 for example, in the case of a vanishing chemical potential, the
 number density
is  Boltzmann suppressed for large fields.

\begin{center}
\section{\sc The Physical Content of $\cL_{eff}$}
\label{physical}
\end{center}
\setcounter{equation}{0}
There are several dimensionful parameters related to $\cL_{eff}$, i.e.
$T,\ \mu,\ m$, and $B$, that can be large or small
compared to each other. We shall discuss some
of these limits which we think are particularly
interesting.
A central feature of a fermion gas is whether it is
degenerate or not, i.e. whether or not the Fermi surface is sharp
on the scale of the Fermi energy. With an external magnetic
field it is also important to compare the smoothness of the
Fermi surface with the spacing of the Landau levels.
A criteria for the  de\ Haas -- van\ Alphen effect is that the distance
 between the Landau levels close to the Fermi surface
is considerably
 larger than
the diffuseness or fluctuations in the Fermi surface due to finite temperature,
electron -- electron interactions, impurities etc.
This can sometimes be achieved even at high $T$ by having large
$\mu$ and $B$.

The effective Lagrangian is here given as a function of
the chemical potential $\mu$. In many situations it is more
natural to consider the expectation value of charge density $Q/V$ as given,
where $Q$ is the total conserved charge. It is calculated from $Q/V =
-e\rho(\mu)$, where
\be
      \rho(\mu)=-\inv{V}\frac{\del F}{\del \mu}
      =\frac{\del \lbmeff}{\del\mu}~~~ ,
\ee
which in the case of  vanishing magnetic field and temperature reduces to
\be
\sqrt{\mu^{2}-m^{2}}=(3\pi^{2}|\rho |)^{1/3}~~~,
\ee
and  $\mu$ has the same sign as $\rho$. For large $B$ field this relation gets
substantial
correction, see e.g. Section~\ref{strong}.
We notice that $\rho$ is equal to the
difference between the electron and positron number densities, that may be
 useful on comparison with condensed matter physics calculations.
  In other situations one may
consider adiabatic changes of $B$, and then keep the entropy fixed, or
the pressure. All these different cases are described by
suitable Legendre transformations of the thermodynamical
potential $F$.\vspace{4ex}

%
\subsection{\sc The de\ Haas -- van\ Alphen effect}
\label{dhva}
At low temperature one may attempt an expansion in $T$ using
Sommerfeld's method \cite{AshcroftM76}. We assume that
$\mu>m$ since for $|\mu|<m$ the thermal contribution is
exponentially suppressed. The Sommerfeld expansion for a
function $H(\omega)$ is
\be
      \int_m^\infty d\omega\, f_F^+(\omega)H(\omega)=
      f_F^+(m)\int_m^\infty d\omega\,H(\omega)+
      \sum_{n=1}^\infty T^n a_n
      \left.\frac{d^{n-1}H(\omega)}{d\omega^{n-1}}\right|_{\omega=\mu}\ ,
\ee
where
\be
      a_n=\int_{-\frac{\mu-m}{T}}^\infty dx\frac{x^n}{n!}
      \left(-\frac{\del}{\del x}\inv{e^x+1}\right)\ ,
\ee
but the odd powers of $T$ are exponentially suppressed. This
formula can be applied to $\lbmeffzero$ and $\lbmeffreg$, but
in $\lbmeffosc$ performing the  derivative inside the summation sign is
not allowed since the sum is not uniformly convergent, and when acting on
the form containing the $\zeta$-function there will obviously be divergences
at discrete points.
This indicates that an expansion in $mT/eB$ is not possible.
Anyway, the $T=0$ part of $\lbmeffosc$ can be calculated,
and if we in particular assume $\{T=0,eB\ll\mu^2-m^2\ll m^2\}$
we get
\be
      \lbmeffone \approx
      \frac{(eB)^2}{12\pi^2}
      \frac{\sqrt{\mu^2-m^2}}{m}
      -\frac{(eB)^{5/2}}{4\pi^4 m}\sum_{n=1}^\infty
      \inv{n^{5/2}}\Biggl[\cos\left(\frac{\pi}{4}
      -n\pi\frac{\mu^2-m^2}{eB}\right)-\inv{\sqrt{2}}\Biggr]\ .
\ee
This is a non-relativistic
limit (in the sense that the kinetic energy is much smaller
than $m$) with a degenerate Fermi sea and a weak external field.

The vacuum correction is in this limit given by
\be
\label{SMALLB}
      {\cal L}_{1} \approx \frac{(eB)^2}{360\pi^2}
      \left(\frac{eB}{m^2}\right)^2\ ,
\ee
so that the finite density correction
\be
 \lbmeffone \approx  \frac{(eB)^2}{12\pi^2}
      \left(\frac{3\pi^2 \rho}{m^3}\right)^{1/3}~~~ ,
\ee
therefore  dominates over ${\cal L}_{1}$ when
\be
      \left(\frac{ \rho}{m^3}\right)^{1/3} \gg
      \inv{30(3\pi^2)^{1/3}}\left(\frac{eB}{m^2}\right)^2\ ,
\ee
or equivalently, in terms of the chemical potential
\be
\frac{\sqrt{\mu^{2}-m^{2}}}{m} \gg \frac{1}{30} \left( \frac{eB}{m^{2}}
 \right)^{2}~~~.
\ee
This is always satisfied in the limit $\{eB\ll\mu^2-m^2\ll m^2\}$.

Even though the $B^2$ dominates over $B^{5/2}$ for small $B$, the
magnetization of the heat bath\footnote{The vacuum contribution to the
magnetization is not included in \Eqref{thermmag} since it is very small for
small $B$.}
gets a larger contribution from $\lbmeffosc$,
\be
\label{thermmag}
      M=M_{reg}+M_{osc}=-\inv{V}\frac{\del F}{\del B}
	=\frac{\del \lbmeff}{\del B}~~~,
\ee
where to the lowest order in the magnetic field
\bea
\label{Mosc}
     M_{osc}&=& \frac{e\sqrt{eB}(\mu^2-m^2)}{4\pi^3 m}
      \sum_{n=1}^\infty \inv{n^{3/2}}
      \sin\left(\frac{\pi}{4}-n\pi\frac{\mu^2-m^2}{eB}\right) \nonumber \\
    &=&-\left(\frac{e}{2m}\right) \frac{\sqrt{2eB}}{\pi^{2}}(\mu^{2}-m^{2})
        \zeta \! \left( -\frac{1}{2},\mbox{ mod}\! \left[ \frac{ \mu^{2}
   -m^{2}}{2eB} \right] \right)~~~,
\eea
and
\be
      M_{reg}=\left(\frac{e}{2m}\right)
      \frac{eB}{3\pi^{2}}\sqrt{\mu^{2}-m^{2}}~~~.
\ee
The $\zeta$-function has its maximal modulus at  $\zeta \!
 \left( -\frac{1}{2},1
 \right) = \zeta \!\left( -\frac{1}{2},0 \right) \approx -0.208$, which implies
that the peak magnetization from the
oscillating term is  larger than  that from the regular term for
\{$eB\lsim 0.78(\mu^2-m^2)$\}, i.e. when the approximations
used here are valid. Defining the magnetic susceptibility as the response in
 the magnetization due to a magnetic field, i.e. $M=\chi B$, as in
Ref.\cite{Isihara91}, we get exact agreement with this reference, but
  not with Refs.\cite{Abrikosov88,Kittel63},
which have an extra factor
 $(-1)^{n}$ in the sum over $n$, that we find only should be present in the
case of spinless bosons.
In Section~\ref{strong} we give an argument why our result
has to be correct.

The oscillatory behaviour as a function of $B$ is well-known as
the de\ Haas -- van\ Alphen effect.
The frequency of this periodic function
agrees with the one derived by Onsager \cite{Onsager52}.
Equation~(\ref{Lbmueff}) describes
the full relativistic generalization of this effect,
and in Section~\ref{astro} we consider some astrophysical
applications where the non-relativistic approximation is
not valid. The distance
between the magnetic field of two
adjacent minima of the magnetization is determined
by
\be
      \abs{\inv{eB_i}-\inv{eB_{i+1}}}= \frac{2\pi}{A}\ ,
\ee
where $A$ is the area of an extremal cross section of the
Fermi sea.

Sometimes (e.g. in Ref.\cite{AshcroftM76})
 the magnetic susceptibility is defined by
\be
      \chi=\frac{\del M}{\del B}\ ,
\ee
but again we find that the sum over $n$ does not converge, and that the form
containing the $\zeta$-function  contains divergences at discrete values
of $B$, and is poorly illuminating.
%
\subsection{\sc Strong B-field}
\label{strong}
In the limit of strong field, $\{eB\gg T^2,m^2,\abs{\mu^2-m^2}\}$,
we can see from \Eqref{partition} that only the lowest Landau level
contribute and $\lbmeff$ goes like a linear function of $eB$.
We shall now reproduce this result from \Eqref{Lbmueff}
and it turns out to be rather non--trivial. The leading $B$
dependence in the first term in \Eqref{Lbmueff} is obtained by
scaling out $eB$ and taking $eB\rightarrow \infty $ in the remainder.
The total contribution is, apart from the thermal integration (see
Appendix B)
\be
      \frac{(eB)^{3/2}}{4\pi^{5/2}}\left[
      \int_0^\infty \frac{dx}{x^{5/2}}(x\coth x -1)
      -\sqrt{\frac{2}{\pi}}
      \sum_{n=1}^\infty\inv{n^{3/2}}\right]~~~,
\ee
but this is actually identically zero. The next subleading
term can be shown to be
\be
\label{laBLett}
      \lbmeffone = \frac{eB}{2\pi^2}\int_{-\infty}^\infty
      d\omega\theta(\omega^2-m^2)f_F(\omega)
      \sqrt{\omega^2-m^2}~~~,
\ee
which is exactly the leading term from \Eqref{partition}. This
calculation shows that the oscillatory term in \Eqref{Lbmueff}
is absolutely necessary to cancel the $B^{3/2}$ term and to give
the correct linear term. Also, notice that the expression  presented here
 for this term
has to be correct, without the extra factor $(-1)^{n}$ of
Refs.\cite{Kittel63,Abrikosov88},
for the $B^{3/2}$ terms to cancel.

In this limit of strong magnetic field the
thermal and density  corrections given above are small compared to
\be
\label{LARGEB}
      {\cal L}_{1} \approx \frac{(eB)^2}{24\pi^2}
      \log\left(\frac{eB}{m^2}\right)\ .
\ee
The vacuum polarization
effects are dominating here, which comes quite naturally, since the
magnetization from  real
thermal particles becomes saturated when all spins are aligned, whereas the
magnetization  from vacuum polarization increases like $B\log B$. This has not
always been recognized in the literature \cite{MillerR84}.

Another issue  when the $B$ field is strong compared to $\mu^2-m^2$ is that the
relation between $\rho$ and $\mu$ is changed.
In fact, we have from \Eqref{laBLett} at $T=0$
\be
      \rho(\mu)\approx\frac{eB}{2\pi^2}\sqrt{\mu^2-m^2}\ .
\ee
The linear dependence on the Fermi momentum
$k_F=\sqrt{\mu^2-m^2}$ can be understood
from the fact that only the lowest Landau level is filled
and therefore the phase space is essentially one-dimensional.
%
\subsection{\sc Weak B-field}
\label{weak}
In Section \ref{dhva} we had an expression for $\lbmeff$
in a weak ($\ll\mu^2-m^2$) field but $T^2$ still smaller
than $eB$. An expansion for $B$ smaller than all other scales
would be desirable but there are some subtleties involved in such an expansion.
The vacuum part can be expanded in a naive way and we get
\be
	\cL_1=- \frac{m^4}{4\pi^2}\sum_{k=1}^\infty
	\left(\frac{eB}{\pi m^2}\right)^{2k+2}(-1)^k\zeta(2k+2)\Gamma(2k)\ .
\ee
This series is not convergent but Borel summable for small $eB/m^2$ so we
expect the first few terms to be a good approximation for weak fields.
Expanding the integrand of $\lbmeffreg$ (see \Eqref{Lbmueff}) in powers of $B$
leads to the same problem after the $s$-integration. Moreover, the
$\omega$-integration becomes  infra--red divergent, for higher order terms. We
cannot even expand the integrand of $\lbmeffosc$ in powers of $B$, but after
repeated
 partial integrations with respect to $\omega$ we obtain
\bea
      \lbmeffosc &=& \frac{m^4}{4\sqrt{2}\pi^{3/2}}\sum_{k=0}^\infty
      \left(\frac{eB}{\pi m^2}\right)^{5/2+k}\zeta(5/2+k)
      (-1)^{[k/2]}  \non \\
 &&\times \left( m^2 \frac{d}{d\omega^2}\right)^k
      \left.\frac{m}{\omega}
      \Bigl(f^+_F(\omega)+f^-_F(\omega)\Bigr)\right|_{\omega=m}\ ,
\label{wBexp}
\eea
where $[k/2]$ is the integral part of $k/2$.
When $|\mu|>m$ the factor with derivatives of $f^\pm_F(\omega)$
at $\omega=m$ contains powers of $m/T$. These factors,
combined with the $B/m^2$ factors, show that we must
have $\{B\ll m^2,T^2\}$ in order for the expansion to be valid.
 For $|\mu|<m$ these terms are exponentially suppressed at
small $T$. We thus see that there is an intricate interplay
between $B$ and $T$ in such a way that when \{$eB\ll T^2$\}
$\lbmeffosc$ is smaller than $\lbmeffreg$, as well as their
derivatives. However, when $\{T^2\ll eB\}$, even though
$\{eB\ll\mu^2-m^2,m^2\}$, the $B$ derivatives of $\lbmeffosc$
are large and show a periodic behaviour as shown in
Section \ref{dhva}.
Also the expansion from \Eqref{dittricheqn} is only asymptotic. In view of the
observations above, especially the half--integer powers of $B$ in
\Eqref{wBexp},  it seems unlikely that the same result can be obtained in
ordinary pertubation theory using diagrammatic techniques.
The vanishing radius of convergence for the expansion
of $\cL_1$, and the same for $\lbmeffreg$, also including the infra-red
divergences, arise due to the fact that we get substantial contributions
to the parameter integrals when we are outside the radius of convergence
for the series expansion of the $\coth(eBs)$,
i.e. for large s, for high order terms.
 We will investigate this,  the  non-analyticity in $B$, and the connection to
ordinary perturbation theory
more carefully in a future project.
Some weak--field results can nevertheless be obtained and, for instance, the
magnetic susceptibility can be computed in the limit
\{$B\rightarrow 0$, $T\ll\mu^2-m^2$\}. It gets contribution
only from $\lbmeffreg$,
\be
      \chi=\lim_{B\rightarrow 0}\frac{\del^2\lbmeffreg}{\del B^2}
      =\frac{e^2}{6\pi^2}\log\left(\frac{|\mu|}{m}+
      \frac{\sqrt{\mu^2-m^2}}{m}\right)\ .
\ee
If we further assume that \{\,$\mu^2-m^2\ll m^2$\} and write
it in terms of the Bohr magneton $\mu_B=e/2m$ and the density of
states at the Fermi surface
$g(\mu)=m\sqrt{\mu^2-m^2} /\pi^2$, we find
\be
      \chi=\frac{2}{3}\mu_B^2 g(\mu) =
      \chi_{Pauli}+\chi_{Landau}\ .
\ee
It coincides with the well-known result \cite{AshcroftM76}  where
$\chi_{Pauli}$ is the Pauli paramagnetic  spin
contribution and $\chi_{Landau}=-\inv{3}\chi_{Pauli}$
is the Landau diamagnetic orbital contribution.
Notice that in this weak field limit the thermal corrections dominate, i.e.
\be
  \lbmeffone \approx \frac{(eB)^{2}}{12\pi^{2}} \frac{\sqrt{\mu^{2}-m^{2}}}{m}
\gg {\cal L}_{1}~~~,
\ee
where ${\cal L}_{1}$ is given by \Eqref{SMALLB}.
%
\subsection{\sc High temperatures}
\label{hight}
At high temperatures one may find an analytical approximation in the limit
$\{T^2\gg m^2\gg eB,\mu=0\}$, where we have that
\be
\label{LbmhighT}
      \lbmeffone \approx \frac{(eB)^2}{24\pi^2}
      \log\left( \frac{T^2}{m^2}\right)\ ,
\ee
and we do not agree with the high temperature and weak field
limit in Ref.\cite{dittrich79}.
(We notice the similarity between \Eqref{LbmhighT} and ${\cal L}_{1}$ for
$eB\gg m^2$ in \Eqref{LARGEB}.) The thermal contribution  $\lbmeffone$ thus
dominates over ${\cal L}_{1}$ as  given by Eq.(\ref{SMALLB}) when
\be
      \frac{T}{m}\gg \exp\left[\inv{30}
      \left(\frac{eB}{m^2}\right)^2\right]
      \approx 1\ ,
\ee
i.e. when the approximations used here are valid.
%
\begin{center}
\section{\sc Some Astrophysical Applications}
\label{astro}
\end{center}
\setcounter{equation}{0}
As mentioned in the Introduction, strong magnetic fields
at finite temperature and density are situations that
are frequently encountered in astrophysical contexts.
We have investigated the possibility of some interesting behaviour
mainly for  white dwarfs, neutron stars and
supernovae since they present the most  extreme conditions
while still being directly observable, in contrast to
e.g. cosmic strings, the existence of which has yet to
be confirmed. We can use the effective action in two ways.
 Either we consider the response of the system to a given
external magnetic field $H$, or we study  the properties
of an isolated system with only the induced magnetic field.
In the first case the free energy is given by
\be
      F=-\cL_1(B)-\lbmeff(B)\ ,
\ee
where $B$ is determined by the mean field equation
\be
\label{indB}
      B=H+M(B)=H+\frac{\del\cL_1}{\del B}+
      \frac{\del\lbmeff}{\del B}\ .
\ee
The magnetization $M(B)$ is thus calculated in the presence of
the microscopic magnetic field $B$. Note that we include both
the contribution from real electrons in the heat bath and
virtual electrons from vacuum polarization.
If we consider the dynamics of the system without any external
field we should add $\cL_0$ to the effective action and determine
stationary values of the field by
\be
\label{stat}
      \frac{\del\cL_{eff}}{\del B}=
      -B+\frac{\del\cL_1}{\del B}+
      \frac{\del\lbmeff}{\del B}=0\ ,
\ee
which, of course, is the same as putting $H=0$ in \Eqref{indB}.
As discussed in
Section \ref{strong} the vacuum contribution is dominant for large fields.
Using the result from \cite{MillerR84} we see that for $T=m$ the thermal
contribution saturates at about $eB=10\,m^2$. At that value of the
magnetic field, the vacuum
contribution is about twice as large as the thermal and cannot be
ignored.
\vspace{5ex}\\
It would be most interesting if we could find astrophysical
objects showing the de Haas -- van Alphen oscillations. The magnitude of
the oscillations might then be large enough to effectively
trap the magnetic field in a local minimum satisfying
\Eqref{stat}. A   candidate for such a system is a
neutron star with a strong $B$ field and a degenerate electron gas.
In order to get de Haas -- van Alphen oscillations as a function of $B$ the
spacing of Landau levels near the Fermi surface need to
be larger than the spreading of the Fermi surface due to finite
temperature. If the $n$-th Landau level is at the Fermi surface,
$E_n=\mu$, then we require $E_{n+1}-E_n\gsim T$.
For $\mu^2\gg eB$, which is the case for neutron stars,
we get the condition
\be
      eB\gsim \mu T\ ~~~.
\label{eq-reldhva}
\ee
According to Appendix C we can even get a more stringent condition in the case
of large chemical potential
\be
      eB\gsim 2 \pi^2  \mu T\ ~~~.
\ee
As a comparision, we find in the non-relativistic case
, instead of \Eqref{eq-reldhva} that
\be
   eB \gsim m T~,~~~\{ eB, \mu^2-m^2 \ll m^2\} ~~~.
\ee
In order to see any oscillations the field must not be so high that all
fermions are in the lowest Landau level, i.e.  integer $n$ above must
be greater than unity, that gives
\be
  (\mu^2-m^2)/2 > eB~~~.
\ee

Approximate values for $eB,T$ and $\mu$ for what we find
the most interesting astrophysical
objects in this context, a supernova; a neutron star; and a white dwarf,
are given in Table~1. According to above, the number in the
last two rows of this table should be greater than unity for
 de Haas -- van Alphen oscillations to appear. That is not the case
in either of the situations.
\begin{table}
  \begin{center}
  \begin{math}
  \begin{array}{||l|c|c|c||} \hline \hline
         &{\rm ~~White~ Dwarf} &  {\rm~~ Neutron~ Star~~}
          &{\rm  ~~Supernova~~} \\ \hline
  \mu/m & 1.02~\cite{freese}&  6 \cdot 10^2 ~\cite{neutronstar}
           & 6 \cdot 10^2 ~ \cite{myller90}  \\ \hline
   T/m & 2 \cdot 10^{-3} ~\cite{freese} & 1 ~\cite{neutronstar}
           & 1 \cdot 10^2 ~\cite{myller90} \\ \hline
  eB/m^2 & 2 \cdot 10^{-6} ~\cite{neutronstar,Chanm92}
     & 2 \cdot 10^{-1}~ \cite{Chanm92}  & 2~ \cite{ginzburg91} \\ \hline \hline
  ( \mu^2 -m^2)/(2eB) & 1 \cdot 10^4 & 2 \cdot 10^6
            & 2 \cdot 10^5 \\ \hline
   eB/(\mu T) & 1 \cdot 10^{-3} & 3 \cdot 10^{-4}
       & 3 \cdot 10^{-5} \\ \hline \hline
  \end{array}
  \end{math}
  \end{center}
  \baselineskip 13pt
  \tabcap{Typical values of  $eB,T$ and $\mu$ for some  astrophysical
objects, and an indication of the possibility for oscillations in the
magnetization. The references are given in brackets.}
  \label{tab-astro}
\end{table}
\begin{table}
  \begin{center}
   \begin{math}
   \begin{array}{||c|c|c|c|c||} \hline \hline
    ~ \cL_0~~ (m^4)~ &~\cL_1~~ (m^4)~ & ~ \lbmeffzero~~ (m^4)~
       & \lbmeffreg~~ (m^4)~&
          ~\lbmeffosc~~(m^4)~  \\ \hline
     -2 \cdot 10^{-2} & 2 \cdot 10^{-6} & 1 \cdot 10^9 & 6 \cdot 10^{-2}
      & 1 \cdot 10^{-3}  \\ \hline \hline
  \end{array}
  \end{math}
  \end{center}
  \baselineskip 13pt
  \tabcap{The different parts of the effective Lagrangian
    for a typical neutron star, in natural units.}
  \label{tab-neutrlag}
\end{table}
\begin{table}
  \begin{center}
   \begin{math}
   \begin{array}{||c|c|c||} \hline \hline
    ~ M_1~~ ( e m^2)~ &~ M_{reg}~~ (e m^2)~ &~ M_{osc}~~ (em^2)~  \\ \hline
        2 \cdot 10^{-5} & 4 \cdot 10^{-2} & 1 \cdot 10^{-2}  \\ \hline \hline
  \end{array}
  \end{math}
  \end{center}
  \baselineskip 13pt
  \tabcap{The different parts of the magnetization
    for a typical neutron star, in natural units.}
  \label{tab-neutrmag}
\end{table}
For a neutron star we have numerically computed the different parts
of the effective Lagrangian, and the corresponding magnetization. The
results are given in Table~2 and Table~3,
   respectively.
The effective Lagrangian is totally dominated by the thermal contribution
in absence of a magnetic field , $ \lbmeffzero $, due to the extreme
chemical potential. We would like to stress
  that there are no oscillations in the so called
oscillating part of the magnetization, $M_{osc}$, in this region of
parameters.
Obviously  we do not expect to see any de Haas -- van Alphen oscillations
unless the neutron star is very cold ($T=\cO (1)$ eV), or if the electron
density is very low
in some region, for example close to the surface, where the field still is
strong.
\vspace{5ex}\\
In order to investigate the behaviour of a relativistic gas of fermions
showing de Haas -- van Alphen oscillations, we have numerically calculated
the effective action, and the magnetization for $\{\mu/m=4\,;\
T/m=0.01\,,\ 0.1\,,\ 1.0\}$. The latter is shown in Fig.~1.
\begin{figure}[t]
  \epsfxsize=15cm
  \epsfbox{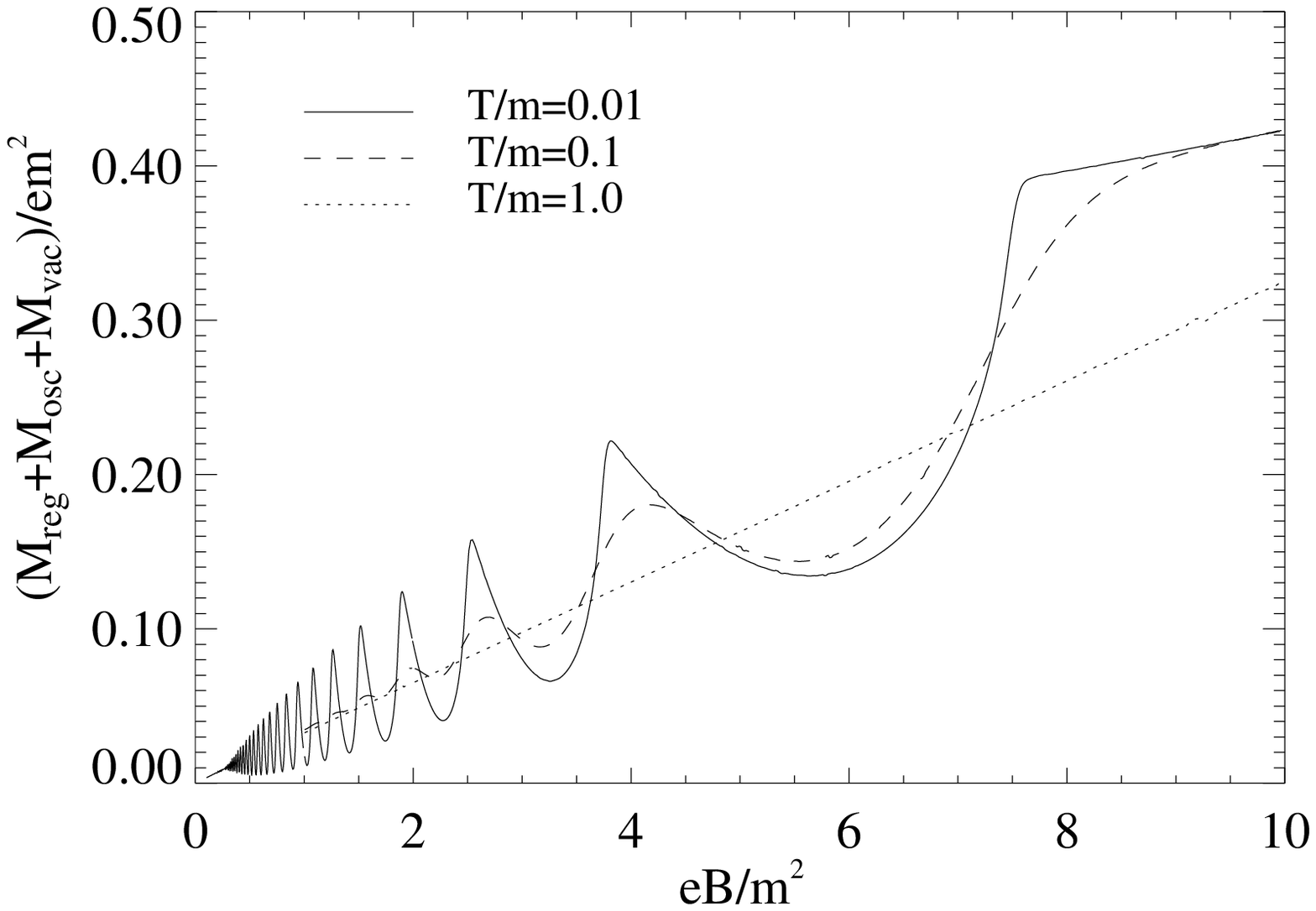}
\baselineskip 13pt
\figcap{The vacuum and thermal contribution to the magnetization showing de
Haas -- van Alphen oscillations
as the temperature is lowered. The chemical potential is $\mu=4\,m$.}
  \nopagebreak
  \label{fig-osc}
\end{figure}
We see clearly how the oscillations disappear as the temperature is raised.
There is also a last oscillation at about $eB\simeq 7m^2$ which occurs when the
second Landau level leaves the Fermi surface. For the values above we do not
find any non--trivial solution to \Eqref{stat} because the tree level $-B$
dominates. It is, in fact, only for a rather limited range of parameters that
$\lbmeffosc$ can give local maxima for the total effective action. As an
example, let us first put $T=0$ since that only enhances the oscillations. Then
we look at small $B$ so that the tree part is small. There is a chance that the
$\sqrt{B}$ term in \Eqref{Mosc} can dominate. Using $\mu\simeq m$ and making
the approximation $\abs{\zeta \!\left( -\frac{1}{2},x \right) }\leq 0.2$, we
get
\be
	\abs{M_{osc}} \lsim 0.2\frac{\sqrt{2}e^{3/2}}{2m\pi^2}
	\sqrt{B}(\mu^2-m^2)
	\simeq 0.005 \sqrt{B} (\mu-m)\ .
\ee
For this term to dominate over $\abs{M_{tree}}=B$ we need $B\lsim 10^{-5}
(\mu-m)^2$ which complicates numerical calculations. Also, since the field is
small the probability of tunneling through the barrier between the maxima is
not very suppressed and it is probably not an efficient way of trapping
magnetic fields.
At very large values of $B$ the vacuum part eventually dominates over the tree
level, but this is just the  Landau ghost and we cannot draw any conclusion
about any instability.

Even if there are no local minima in $-B^2/2-\cL_1-\lbmeff$, there
may be intervals in $B$ where $-\cL_1-\lbmeff$ is concave, i.e. where the
susceptibility is positive.
Domains with different magnetization could then be formed
in presence of an external field, just like in
some solid state materials \cite{Abrikosov88}.

\begin{center}
\section{\sc The Effective QED Coupling}
\label{coupling}
\end{center}
\setcounter{equation}{0}
 The charge renormalization given by \Eqref{chargeren} also leads to the weak
coupling
expansion of the QED $\beta$-function, i.e.
\be
\lambda\frac{d}{d\lambda}\alpha(\lambda) = \beta (\alpha(\lambda)) =
\frac{2}{3\pi}\alpha^{2}(\lambda) + {\cal O}(\alpha^{3}(\lambda))~~~,
\label{BETAF}
\ee
where $\lambda$ is a momentum scale factor.
We notice that due to the scale invariance of $eB$, we
 can also define an effective coupling constant from $\cL_{eff}$  as
\cite{Schwinger51,cos88}
\be
  - \frac{1}{e^2(eB,\mu,T)}=\frac{1}{eB} \frac{ \del \cL_{eff}}{\del (eB)}~~~,
\ee
that gives for the electromagnetic fine structure constant
 $\alpha(eB,\mu,T) \equiv e^{2}(eB,\mu,T)/4\pi$
\be
\label{effalp}
\frac{1}{\alpha(T,\mu,B)}=\inv{\alpha}
-\frac{1}{\alpha B}\frac{\del( \cL_{1} +\cL_1^{\beta,\mu})}{\del B}\ ,
\ee
in analogy with the definition of the renormalized
coupling in the vacuum sector
in connection with \Eqref{BETAF}. Special care has
to be taken when evaluating the
derivative
of the oscillating term in Eq.(\ref{Lbmueff}).
In the limit when $eB=0$, we obtain the effective coupling
$\alpha(T,\mu) = \alpha(T,\mu,B=0)$ given by
\be
\label{AlphaTmu}
\frac{1}{\alpha(T,\mu)}= \frac{1}{\alpha } - \frac{2}{3\pi}\int
_{-\infty}^{\infty}d\omega
\frac{\theta(\omega^{2}-m^{2})}{
\sqrt{\omega^{2}-m^{2}}}f_{F}(\omega)~~~.
\ee
When $T=0$, we therefore get an effective coupling
$\alpha(\mu) = \alpha(T=0,\mu)$ such that
\be
\label{Alphamu}
\frac{1}{\alpha(\mu)}= \frac{1}{\alpha }
-\frac{2}{3\pi}\log \left( \frac{|\mu|}{m} + \sqrt{\frac{\mu ^{2}}{m^{2}} -
1}~\right)\ .
\ee
In the limit $\mu=0$, we find the following asymptotic behaviour
of the corresponding effective coupling $\alpha(T) = \alpha(T,\mu = 0)$,
\be
\label{AlphaT}
\frac{1}{\alpha(T)} = \frac{1}{\alpha } -\frac{4}{3\pi}
\int
_{\beta m}^{\infty}
\frac{dx}{
\sqrt{x^{2}-(\beta m)^{2}}}\frac{1}{e^{x} + 1}
\approx \frac{1}{\alpha }
-\frac{2}{3\pi}\log \left(\frac{T}{m}\right)~~~,
\ee
for $T \gg m $.
It is now clear that (only) for $\mu \gg m$ and $T \gg m$ the
effective couplings
$\alpha (\mu)$ and $\alpha (T)$ are solutions to the renormalization group
equation~(\ref{BETAF}) when $\lambda $ is identified with $\mu$ and $T$
respectively
(see in this context e.g. Refs.\cite{Morley79,rojas92}).
We also note that Eq.(\ref{LARGEB}) leads to an effective
coupling $\alpha(B) = \alpha(T=0,\mu = 0,B)$
with an asymptotic behaviour
\be
\label{AlphaB}
\frac{1}{\alpha(B)} \approx \frac{1}{\alpha }
-\frac{2}{3\pi}\log \left(\frac{\sqrt{eB}}{m}\right)~~~,
\ee
that also satisfies the renormalization group equation \Eqref{BETAF}.
The effective coupling defined in \Eqref{effalp}  can also be extracted
from the residue of the thermal Debye-screened photon propagator
(see Ref.\cite{Morley79}).

The effective couplings as given in Eqs.(\ref{Alphamu}),~(\ref{AlphaT}) and
(\ref{AlphaB}) can be interpreted as follows. If we use the lowest order
 $\beta$-function in \Eqref{BETAF}, then the scale dependent coupling
is given by
\be
\frac{1}{\alpha(\lambda)}=\frac{1}{\alpha}- \frac{1}{3\pi}\log \left(
\frac{\lambda^{2}}{m^{2}} \right)~~~.
\ee
Then we can write
\be
\label{eq:running}
\frac{1}{\alpha(x)}\approx  \frac{1}{\alpha(\lambda)} -\frac{2}{3\pi}
\log \left(\frac{x}{\lambda} \right)~~~,
\ee
where $x=\mu,T$ or $ \sqrt{eB}$. If $\lambda$ is identified with any of
 these scales, we can in each such case write
\be
\cL_{eff}=-\frac{1}{2} \frac{(eB)^2}{e^{2}(x)} +\lbmeffzero~~~,
\ee
when $x \gg ( m \mbox{ and any other scale of dimension energy} )$.

In terms of the effective fine-structure constant, and in the case of small
chemical potentials,
so that $|\mu| < m$, we obtain
\be
      \alpha (eB,T,\mu)
      = \frac{\alpha}{1 - \alpha X(eB,T,\mu)}~~~,
\ee
 where we have defined the functions $X(eB,T,\mu) = X_{1}(eB) +
X_{2}(eB,T,\mu)$,
\be
      X_{1}(eB)=\frac{1}{2\pi} \int_{0}^{\infty} \frac{dx}{x}
      \exp(-x\frac{m^2}{eB}) \left[\frac{1}{\sinh ^2(x)}
      - \frac{\coth(x)}{x} +
      \frac{2}{3}\right]~~~,
\ee
and
\bea
\label{eq:xtwo}
 X_{2}(eB,T,\mu) &=&
   \frac{1}{2\pi}\sum _{l=1}^{\infty}(-1)^{l}
      \int _{0}^{\infty} \frac{dx}{x}
      \exp\left( -\frac{\beta ^{2} l^{2}}{4x} -m^{2}x \right) \nonumber \\
  && \times\left[\frac{1}{\sinh ^2(eBx)} - \frac{\coth(eBx)}{eBx}\right]
      \cosh(\beta l\mu)
        ~~~.
\eea
The function $X_{1}(eB)$ has the following expansions
\be
X_{1}(eB) = \frac{2}{45\pi} \left(\frac{eB}{m^{2}}\right)^{2} + {\cal
O}\left(\left( \frac{eB}{m^{2}}\right) ^{4}\right)
\ee
if $eB \ll m^{2}$ and
\be
X_{1}(eB) = \frac{1}{3\pi}\log \left( \frac{eB}{m^{2}}\right) \left(
1 + \frac{3}{2} \frac{m^{2}}{eB}\right) + {\cal O}\left(\frac{m^{2}}{eB}\right)
\ee
if $eB \gg m^{2}$.
In the case of a vanishing chemical potential we can in \Eqref{eq:xtwo}
identify a
 $\vartheta_{4}$-function, given as
\be
\vartheta_{4}[z,q]=1+2 \sum_{n=1}^{\infty}(-1)^{n} q^{n^{2}} \cos(2nz)~~~,
\ee
and write
\bea
	&& X_{2}(eB,T) = \frac{1}{2\pi}
	\int _{0}^{\infty} \frac{dx}{x} \exp(-x\frac{m^2}{eB})
  	\nonumber \\
  	&& \times
	\left\{1-\vartheta_{4}
	\left[0,\exp\left(-\frac{eB\beta^{2}}{4x}\right)\right]
	\right\}
	\left[  \frac{\coth(x)}{x} -\frac{1}{\sinh ^2(x)} \right]~~~.
\eea
If $eB \ll m^{2}$, we can write
\be
	X_{2}(eB,T) = \frac{4}{3\pi}\sum _{l=1}^{\infty} (-1)^{l+1}
	\left(
	K_{0}(\beta m l) - (\beta m l)^{2}K_{2}(\beta ml)
	{\cal O}
	\left[\left(\frac{eB}{m^{2}}\right)^{2}\right]  \right)~~~.
 \ee
For  $T\gg
m$ we can use
\be
	\sum_{l=1}^\infty K_0(xl)(-1)^{l+1} \rightarrow
	-\inv{2}\log x\ ;\quad x\rightarrow 0~~~ ,
\ee
to find that it leads to a $\log (T/m)$ dependence with the correct prefactor
in accordance with \Eqref{eq:running}. (The approximation of keeping only the
$l=1$, as in Ref.\cite{rojas92},  excludes the factor $1/2$, and thus
 is not correct.) In general we have that
\bea
	&& X_{1}(eB) + X_{2}(eB,T) = \frac{1}{2\pi}
	\int _{0}^{\infty} \frac{dx}{x} \exp(-x\frac{m^2}{eB})
        \nonumber \\
  	&&
	\times\left\{ \frac{2}{3}
	-\vartheta_{4}\left[0,\exp\left(-\frac{eB\beta^{2}}{4x}
    \right)\right] \times
	\left[  \frac{\coth(x)}{x} -\frac{1}{\sinh ^2(x)} \right]\right\}~~~.
\eea

\begin{center}
\section{\sc Discussion and final remarks}
\label{concl}
\end{center}
\setcounter{equation}{0}
\subsection{\sc Inclusion of interparticle interactions}
In our one-loop treatment of the effective action we have not
included interactions between   electrons. The interaction energy between the
particles increases with $T$ and $\mu$ since the density increases, but so does
kinetic energy. For a degenerate
electron gas with large chemical potential the kinetic energy
dominates over the potential energy for electrons close to the
Fermi surface. However, not all electrons have large kinetic
energy and corrections from interactions have to be considered for electrons
with low momenta.
The self-energy correction for fermions at high temperature
and density, but zero external field, has been computed in e.g.
Refs.\cite{Petitgirard92,AltherrK92} (in Ref.\cite{AltherrK92}
only massless fermions were considered but it gives an indication
of the correction, especially in view of the result in
Ref.\cite{Petitgirard92}). There appear some completely
new collective phenomena, such as hole excitations
\cite{Weldon89}, which are
not taken into account in this paper.
For the particle excitations the dispersion relation can be
approximated  by an ordinary massive particle
provided the mass is replaced by an effective $T$ and $\mu$
dependent mass \cite{Petitgirard92}
\be
\label{mp}
      m_p=\frac{\sqrt{m_e^2+4M^2}+m_e}{2}\ ,
\ee
where $M$ is the thermally induced mass which in the case of
QED is
\bea
      M^2=\frac{e^2 \mu^2}{8\pi^2}
      &;&
      \quad T=0,\ \mu\neq 0\ ,\non\\
      M^2=\frac{e^2T^2}{8}
      &;&
      \quad T\neq 0 ,\ \mu=0\ ,
\eea
at least if $T \gg m$ and $ \mu \gg m$.
 The hole excitation has
a more peculiar dispersion relation but its spectral
weight is on the other hand lower. It is difficult
to make any quantitative estimates of the importance
of self-energy corrections. We do not, however, expect that phenomena like the
de Haas -- van Alphen oscillations should be altered since it depends on the
electrons at the Fermi surface.
\subsection{\sc Further developments}
There are some extensions of our work that may be of
physical importance. First, we can consider the self-energy
correction of an electron in presence of an external $B$ field.
{}From that the anomalous magnetic moment can be extracted
and compared with previous calculations for small $B$ field,
where there appears some problems of analyticity in the
external photon momentum at finite density \cite{pebss91}.
The self-energy is also important for the higher loop
corrections of the effective action as discussed above.
The QED radiative corrections
could
effect the
electroweak transition rates, relevant for the Big-Bang primordial
nucleosynthesis \cite{ChengST93}.
The photon polarization tensor should also be calculated, and in particular its
imaginary part which is related to the decay into an $e^+e^-$--pair.
Also the three-photon vertex is interesting since it does not exist in absence
of the external field. Such photon splitting processes have been considered
earlier in vacuum \cite{AdlerBCR70,BialynickaB70,BrezinI71,PapayanR72} and it
would be interesting to study the correction from a thermal environment.

The physically more complicated case of a constant (or slowly varying) $E$
field
is equally interesting. A plasma does not stay in equilibrium
since the $E$ field  gets screened and the physical
picture is very different from the one discussed in this paper.
Yet another generalization would be to expand a
non-constant field in powers of the derivative.
Such an expansion has been studied in Ref.\cite{Hauknes84}
at zero temperature.
\subsection{\sc Conclusion}
The main objective of this paper
has been to establish the correct form of the one-loop
QED effective action at finite temperature and density
to all orders in a constant external magnetic field, and the
result differs from earlier attempts. From the form of
$\lbmeff$ presented in \Eqref{Lbmueff} we have checked several
limits that can be understood from a physical point of view.
A great advantage  with our expression for $\lbmeff$ is that
the thermal distribution function $f_F(\omega)$ occurs explicitly.
This means that it is easy to study other thermal situations
by simply replacing $f_F(\omega)$ with some other (non-equilibrium)
distribution (see e.g. Ref.\cite{ElmforsEV93c}).
The importance of the thermal correction depends on the value
of $B$, $T$ and $\mu$. In some physically interesting cases
they may be large compared to $m$ but  often of the same
order of magnitude, which makes it difficult to obtain
analytical approximations. It is, however, possible
to use \Eqref{Lbmueff}, or the expressions in Appendix C, for numerical
calculations.

Even though the correction to the free energy may be small
compared to the value without the external field there
are other quantities that are effected by the presence of
the heat bath. For instance, the magnetization of a degenerate
Fermi sea shows the de Haas -- van Alphen effect.
We found, however, that  for a neutron star this effect does not show up in
spite of the extreme degeneracy and magnetic field. The reason is the
relativistic form of the energy spectrum which suppresses the oscillations at
a large chemical potential. We also briefly discussed  the  importance of
including the vacuum contribution to the magnetization when the $B$ field is
comparable to $m^2/e$.

We have, furthermore, calculated an effective coupling constant
defined from the  derivative of $\lbmeff$ with respect to $B$.
It satisfies asymptotically a naive zero temperature
renormalization group
equation where the renormalization scale is replaced by
$T$, $\mu$ or $\sqrt{eB}$.

%
\vspace{3mm}
\begin{center}
{\bf ACKNOWLEDGMENT}
\end{center}
\vspace{3mm}
One of the authors (B.-S.~S.) would like to thank
John Ellis for
the hospitality of the Theory Division at CERN where some of this work was
initiated and NFR for providing the financial support.
P.~E. wants to thank C.~Pethick for discussions about
neutron stars. It is a pleasure to thank the organizers of the 3rd Workshop
on Thermal Field Theories, 1993, and in particular R. Kobes and G. Kunstatter,
for
providing a stimulating atmosphere during which parts of the present work were
finalized.
\vspace{3mm}
%
\renewcommand{\thesection}{A}
\setcounter{section}{1}
\setcounter{equation}{0}
\begin{center}
{\Large {\sc Appendix A}}
\end{center}
In this appendix we give some details of how to
calculate the effective action in \Eqref{Lbmueff}.
First we show that \Eqref{dittricheqn}
is equal to \Eqref{Lbmueff}.
To do that we start with a Poisson resummation in $l$ using
\be
\label{Poisson}
      \sum_{l=1}^\infty(-1)^l\exp(-\frac{l^2}{4a})=
      \sqrt{4\pi a}\sum_{l=0}^\infty \exp(-a\pi^2(2l+1)^2)
      -\inv{2}\ ,
\ee
and rewrite the sum over $l$ as a contour integral by means
of the formula
\be
      \sum_{l=0}^\infty f\bigl(\frac{\pi}{\beta}(2n+l)\bigr)=
      \frac{\beta}{2\pi}\int_C\frac{d\omega\,f(\omega)}
      {e^{-i\beta\omega}+1}\ .
\ee
The integration contour $C$ is chosen to go from
$\infty+i\epsilon$ to $\epsilon$ in the upper half plane
and back to $\infty-i\epsilon$ in the lower half plane
(i.e. $\omega\in\{\infty+i\epsilon\rightarrow
\epsilon\rightarrow\infty-i\epsilon\}$), without encircling the origin.
In this way all the poles on the positive real axis are
encircled.

We would now like to deform the $\omega$-integral to the
imaginary axis and the $s$-integral to the negative real axis.
This is not   straightforward since there are poles on the imaginary
$s$-axis and the section at infinity has to bo chosen to give a vanishing
contribution. It is,
therefore, necessary to divide the integral into several pieces and
to do the deformation for each piece separately. Let us start with
the part where $\omega$ is in the upper half plane. Then the $s$-contour
can be deformed to the negative imaginary axis, but to the right of the
poles. After that the $\omega$ contour is deformed to the positive
imaginary axis. Finally, for $\abs{\omega}>m$ we further continue
the $s$-integral to the negative real axis and pick up the poles on the
negative imaginary axis, while for $|\omega|<m$ we deform the
$s$-contour back to the positive real axis.

The whole procedure can
be repeated for $\omega$ below the real axis, reflecting all
deformations around the real axis. To get the correct convergence for
the $\omega$-contour deformation, the constant $-1/2$ in \Eqref{Poisson}
should be associated with the $\omega$ in the lower half plane.
After summing the pieces there is only a contribution from
$|\omega|>m$, as expected, and it consists of an $s$-integral
and a sum over the residues of the poles.

In the deformations above we have been careful with the convergence
for large $|s|$ and $|\omega|$, but we have said nothing
about the possible singularity at $s=0$. One way of dealing with
that is to multiply the expression with $s^\nu$ and perform
the integration for such a $\nu$ that there is no
divergence at $s=0$, and to do the analytic continuation at the end.

Equation (\ref{Lbmueff}) can also be obtained from the thermal
propagator in \Eqref{btprop} by representing the $\delta$-function
as
\be
      2\pi i\delta(x)=i\im\inv{x-i\epsilon}=
      i\im\int_0^\infty ds\,e^{-i(x-i\epsilon)}\ .
\ee
Then the $k_y$ and $k_z$ integrations can be carried out
(using \Eqref{Iid} as well). The summation over $n$ is just a
geometric series but it is not absolutely convergent so we
sum only to a finite $N$ and take
the limit $N\rightarrow\infty$ at the end. This gives
\bea
      \Tr S_F^{\beta,\mu}(x;x|m)&=&  \lim_{N\rightarrow\infty}
      i\,\frac{mB}{\pi^{3/2}}
      \,\im\int_{-\infty}^{\infty}\frac{d\omega}{2\pi}f_{F}(\omega)
      \int_0^\infty\frac{ds}{s^{1/2-\nu}}e^{i\frac{3\pi}{4}}
      e^{-is(\omega^2-m^2-i\epsilon)}   \non \\
   && \times\left[\frac{1+e^{i2sB}}{1-e^{i2sB}}-
      \frac{2 e^{i2NsB}}{1-e^{i2sB}}\right]~~~,
\label{SFbmu}
\eea
where we also have introduced the dimensional regularization $\nu$ in $4-2\nu$
dimensions, and we are to analytically continue to $\nu=0$ in the end.
Keeping $\nu$   large enough that the integral is absolutely convergent,
the expression above can easily be integrated with respect to
$m$ to yield $\lbmeff$.     To be more precise, there is an
integration constant from the lower limit in
\be
\label{mint}
    i  \int_{m_0}^{m}dm'\Tr\,S_F^{\beta,\mu}(x;x|m') =
      \lbmeff(m)-\lbmeff(m_0)\ ~~~.
\ee
We expect that in the limit $m\rightarrow\infty$ the
thermal part of the effective action is zero since an
infinitely massive particle has zero Boltzmann weight.
Therefore we let $m_0 \rightarrow \infty $ and thereby
 put the integration constant $\lbmeff(m_0)$
to zero.

The poles in the last factor in \Eqref{SFbmu} cancel for finite
$N$, and we cannot let $N\rightarrow\infty$ in a naive
way before deforming the $s$ integration contour to the
imaginary axis. The two terms have to be treated separately
so we must choose an integration contour for $s$ slightly
above or below the real axis.
Since, according to the discussion above,  $\lbmeff(m \rightarrow \infty)=0$
we see that the the original contour must be chosen slightly above the real
axis.
 Depending on the sign
of $\omega^2-m^2$ (or $\omega^2-m^2-2eB(N-1)$ in the the second
term) we deform the $s$-contour to either the positive or
negative imaginary axis. In one of the cases we get a
contribution from the poles.
After deforming the contours we take the $N\rightarrow\infty$
limit and also take the limit $\nu \rightarrow 0$ what concerns taking the
imaginary part, in order to get a more apparent expression, but we still need
to keep $\nu >0$ to have the integration over $s$ finite, with the result
\bea
      \hspace{-3ex}\lbmeff &=&
      \int_{-\infty}^\infty d\omega
      \theta(\omega^2-m^2)f_F(\omega)
      \Biggl[\inv{4\pi^{5/2}}\int_0^\infty
      \frac{ds}{s^{5/2-\nu}}e^{-s(\omega^2-m^2)}
      seB\coth (seB) \Biggr]\non\\
 &-&
\label{eqleffnu}
      \int_{-\infty}^\infty d\omega
      \theta(\omega^2-m^2)f_F(\omega)
      \Biggl[
      \inv{2\pi^3}\sum_{n=1}^\infty \left(\frac{eB}{n}\right)^{3/2}
      \sin \! \left(\frac{\pi}{4}-\frac{\pi n}{eB}
      (\omega^2-m^2)\right) \Biggr]\ .
\eea
 Actually we must
have $\nu >3/2$, i.e. less then one dimension, but we may just consider it as
an analytical continuation in $\nu$, in order to be able to change the
order of integration.
If we now take the limit $B \rightarrow 0$, we get
\bea
  \lbmeffzero&=& \frac{1}{4 \pi^{5/2}}  \int_{-\infty}^\infty d\omega
      \theta(\omega^2-m^2)f_F(\omega) \int_0^\infty ds\, s^{\nu-5/2}
  e^{-s(\omega^2-m^2)} \nonumber \\
&=&
 \frac{1}{4 \pi^{5/2}}  \int_{-\infty}^\infty d\omega
      \theta(\omega^2-m^2)f_F(\omega) (\omega^2 -m^2 )^{3/2-\nu}
   \Gamma(\nu-3/2)~~~.
\eea
We may now take the limit $\nu \rightarrow 0$ to get  \Eqref{lbmeffzero},
and after subtraction of this term we may also let $\nu$ vanish in
\Eqref{eqleffnu} and get \Eqref{Lbmueff}.
\renewcommand{\thesection}{B}
\setcounter{section}{1}
\setcounter{equation}{0}
\begin{center}
{\Large {\sc Appendix B}}
\end{center}
In the limit of very strong fields $ \{eB \gg T^{2},m^{2}, |\mu^{2}-m^{2}| \}$,
the first term in \Eqref{Lbmueff} can be written as
\be
\lbmeffreg=\int_{-\infty}^{\infty}d\omega \theta(\omega^{2}-m^{2})
f_{F}(\omega) \frac{(eB)^{3/2}}{4\pi^{5/2}} \int_{0}^{\infty}
\frac{ds}{s^{5/2}}(s \coth s -1)~~~.
\ee
Similarly we find in this limit
\be
\lbmeffosc=-\int_{-\infty}^{\infty}d\omega \theta(\omega^{2}-m^{2})
f_{F}(\omega) \frac{(eB)^{3/2}}{2\sqrt{2}\pi^{3}} \zeta(3/2)~~~,
\ee
where $\zeta$ is the Riemann  $\zeta$-function.
It can be shown by residue calculations that
\be
 \int_{0}^{\infty}\frac{ds}{s^{5/2}}(s \coth s -1)=\sqrt{\frac{2}{\pi}}
 \zeta(3/2)~~~,
\ee
so that the $O(B^{3/2})$ terms cancel in this limit. In order to extract
the next term in the strong field expansion of $\lbmeffone$, we consider the
expression entering in the $\omega$ integral in $\lbmeffone$, expanded for
 large $B$, i.e.
\bea
	\frac{(eB)^{3/2}}{4\pi^{5/2}} \left\{  \int_{0}^{\infty}
	\frac{ds}{s^{5/2}}
	(\exp[-\frac{s}{eB}(\omega^{2}-m^{2})]-1)(s \coth s -1)
	\right. && \nonumber \\
 	\left. -\frac{2}{\sqrt{\pi}}
	\sum_{n=1}^{\infty}\frac{1}{n^{3/2}} \left(
	\sin[\frac{\pi}{4}-\frac{n\pi}{eB}
	(\omega^{2}-m^{2})]-\sin\frac{\pi}{4} \right)
	\right\}&&~~~.
\eea
If we now use the cancellations depicted above, and the fact that the
sum converges towards an integral in the limit $B \rightarrow \infty$, the
expression above may be written as
\bea
 	&&\frac{(eB)^{3/2}}{4\pi^{5/2}} \left\{  \int_{0}^{\infty}
	\frac{ds}{s^{3/2}} (\exp[-\frac{s}{eB}(\omega^{2}-m^{2})]) \right.
   \nonumber \\
&&
	-\left. \frac{1}{\sqrt{\pi B}} \int_{0}^{\infty} \frac{dx}{x^{3/2}}
	\left( \sin[\frac{\pi}{4}-x\pi
	(\omega^{2}-m^{2})]- \frac{1}{\sqrt{2}} \right)
	\right\}~~~.
\eea
By performing the integrals in this expression, we find the following leading
 contribution
\be
\lbmeffone=\frac{eB}{2\pi^{2}}
\int_{-\infty}^{\infty}d\omega \theta(\omega^{2}-m^{2})
f_{F}(\omega) \sqrt{\omega^{2}-m^{2}}~~~.
\ee
\pagebreak
\renewcommand{\thesection}{C}
\setcounter{section}{1}
\setcounter{equation}{0}
\begin{center}
{\Large {\sc Appendix C}}
\end{center}
In the case of large chemical potentials, e.g. in a neutron star,
when \mbox{ $ (\mu^2-m^2)/2eB \gg 1$}, the form for
$\lbmeffosc$ given in \Eqref{eqovan} is difficult to handle due to the rapid
oscillations in the $\zeta$-function. Let us instead start from
\Eqref{Lbmueff}, and rewrite it as
\be
  \lbmeffosc= \frac{m^4}{2 \pi^3} \left( \frac{eB}{m^2} \right)^{3/2}
  \sum_{n=1}^{\infty}n^{-3/2} \im \left\{ \exp\left[-i\left(
\frac{\pi}{4} +\frac{\pi n}{eB/m^2} \right) \right] I_{n} \right\}~~~,
\label{eqosc}
\ee
where we have defined
\be
  I_{n} \equiv \int_{1}^{\infty} dx \frac{ \exp\left(i
\frac{\pi n}{eB/m^2} x^2 \right) }{1+\exp[m\beta(x-\mu/m)]}~~~.
\ee
Since the exponential function here is oscillating rapidly and we desire
a rapidly decreasing function instead, we close the
contour with a circular section at infinity, a straight line from the origin
to infinity with complex argument $\pi/4$, and the small section from
the origin to $x=1$, and use Cauchy's theorem to get
\be
  I_{n}=e^{i\pi/4} \int_{0}^{\infty}dx  \frac{  \exp\left(-
\frac{\pi n}{eB/m^2} x^2 \right) }{1+\exp[m\beta(e^{i\pi/4} x-\mu/m)]} -
 \int_{0}^{1} dx  \frac{ \exp\left(i
\frac{\pi n}{eB/m^2} x^2 \right) }{1+\exp[m\beta(x-\mu/m)]} + I_{n}^{poles}~~~.
\ee
The contribution from the residues at the poles is
\be
  I_{n}^{poles}=-2\pi i \frac{T}{m} \exp\left[ -2\pi^{2}\frac{\mu T}{eB}
  +i \pi n \frac{\mu^2}{eB} \right]
 \sum_{\nu=0}^{\nu_{max}} \exp \left[ -2 \pi^{2}n \frac{\mu T}{eB} 2 \nu -
   i \pi^3 \frac{T^2}{eB}(2 \nu +1)^2  \right]~~~,
\ee
where we have defined $\nu_{max}$
 as the number of poles encircled by the contour
\be
   \nu_{max}=\mbox{ int}\left[ \frac{\mu}{2\pi T}-\frac{1}{2}\right]~~~.
\ee
In the case of large chemical potential compared to the temperature and
the square root of the magnetic field, we may
assume the thermal distributions to be unity,
  and perform the integrals with the result
\be
\label{eq:davidsin}
  I_{n}= e^{i \pi/4}\, \frac{1}{2} \sqrt{ \frac{eB/m^2}{n}}
\left( 1- \mbox{erf}\left[ \sqrt{ \frac{n}{eB/m^2}} e^{-i\pi/4} \right] \right)
+ I_{n}^{poles} + O[e^{-\beta( \mu -\sqrt{\frac{eB}{2\pi}})}]~~~.
\ee
It turns out that the phase from one
 minus the error function in \Eqref{eq:davidsin} cancels the phase
from $ \exp\left[-i\left(
\frac{\pi}{4} +\frac{\pi n}{eB/m^2} \right) \right]$  in \Eqref{eqosc}, when
taking the imaginary part. The oscillations are thus only originating from
the residues at the poles, that all have $\re[\omega]=\mu$, i.e. they are lying
at the Fermi surface. Also, notice that the contribution from these poles is
exponentially suppressed as $ \exp\left[ -2\pi^{2}\frac{\mu T}{eB} \right]$,
in agreement with the general discussion on de Haas -- van Alphen oscillations
in Section~\ref{astro} .
%

%
\end{document}